# Digging deeper: deep joint species distribution modeling reveals environmental drivers of Earthworm Communities

**Authors:**

Si-moussi Sara[1*],Thuiller Wilfried[1],Galbrun Esther[2], Decaëns Thibaud[3],Gérard Sylvain[4], Marchán Daniel F.[5],Marsden Claire[4],Capowiez Yvan[4,6], Hedde Mickaël[4*]

**Affiliations:**

1. Univ. Grenoble Alpes, Univ. Savoie Mont Blanc, CNRS, LECA, Grenoble, France
2. School of Computing, University of Eastern Finland, Kuopio, Finland
3. CEFE, Univ. Montpellier, CNRS, EPHE, IRD, Montpellier, France
4. Eco&Sols, INRAE, IRD, CIRAD, Institut Agro, Montpellier, France
5. Department of Biodiversity, Ecology and Evolution, Faculty of Biology, Universidad Complutense de Madrid, Spain
6. Emmah, INRAE, Univ Avignon, France

* Corresponding author: Sara.si-moussi@univ-grenoble-alpes.fr

# Abstract

### Background and aims

Earthworms are key drivers of soil function, influencing organic matter turnover, nutrient cycling, and soil structure. Understanding the environmental controls on their distribution is essential for predicting the impacts of land use and climate change on soil ecosystems. While local studies have identified abiotic drivers of earthworm communities, broad-scale spatial patterns remain underexplored.

### Methods

We developed a multi-species, multi-task deep latent variable model to jointly predict the distribution of 77 earthworm species across metropolitan France, using historical (1960–1970) and contemporary (1990–2020) records. The model integrates climate, soil, and land cover variables to estimate habitat suitability. We applied SHapley Additive exPlanations (SHAP) to identify key environmental drivers and used species clustering to reveal ecological response groups.

### Results

The joint model achieved high predictive performance (TSS > 0.7) and improved predictions for rare species compared to traditional species distribution models. Shared feature extraction across species allowed for more robust identification of common and contrasting environmental responses. Precipitation variability, temperature seasonality, and land cover emerged as dominant predictors of earthworm distribution. Species clustering revealed distinct ecological strategies tied to climatic and land use gradients.

### Conclusions

Our study advances both the methodological and ecological understanding of soil biodiversity. We demonstrate the utility of interpretable deep learning approaches for large-scale soil fauna modeling and provide new insights into earthworm habitat specialization. These findings support improved soil biodiversity monitoring and conservation planning in the face of global environmental change.

# 1 Introduction

Earthworms are among the most ecologically significant invertebrates in terrestrial ecosystems, contributing substantially to overall animal biomass and playing an essential role in soil functioning (Blouin et al., 2013) . These organisms are pivotal in driving key soil processes such as organic matter decomposition, nutrient cycling, and the formation of soil structure (Vidal et al., 2023) . Through their burrowing activities, earthworms increase soil porosity, enhance water infiltration, and promote plant growth, making them crucial for ecosystem productivity and soil health (Arrázola-Vásquez et al., 2022; Jouquet et al., 2006) . Additionally, as key components of soil food webs, earthworms connect various soil organisms, influencing nutrient flows and energy transfer across trophic levels below and above ground (Goury et al., 2025; Wan et al., 2025). Understanding their spatial distributions is therefore crucial for predicting ecosystem responses to environmental change such as climate shifts or land use transformations.

While local-scale studies have identified several abiotic drivers shaping earthworm distributions (Decaëns et al., 2008; Eggleton et al., 2009; Eisenhauer et al., 2014; Pelosi et al., 2014), large spatial patterns remain less understood (Mathieu and Jonathan Davies, 2014; Phillips et al., 2019; Zeiss et al., 2022). This gap in knowledge is particularly critical in the context of the proposed EU Directive on Soil Monitoring and Resilience, which emphasizes the urgent need for robust indicators of soil health and biodiversity at large scale (European Commission, 2023). Improving our understanding of soil organism responses to environmental changes is therefore essential. While traditional earthworm species classifications based on morphological traits (Bouché, 1977) have long guided ecological interpretations of species' occurrences, they can be subjective and inconsistently applied (Bottinelli and Capowiez, 2021). A more objective approach could involve the modeling of species responses to environmental gradients, enabling data-driven ecological classifications and improved predictive accuracy.

Single-species species distribution models (SDMs) are commonly used to predict species distributions based on environmental predictors (Guisan et al., 2017). However, many earthworm species are rather rare, often with sparse observations, making single-species SDMs prone to overfitting or underfitting. Also, in many single-species modeling workflows, predictions for each species are later "stacked" to infer community composition (Grenié et al., 2020). This process can lead to overestimation of species richness when summed across all single-species predictions (Calabrese et al., 2014; Deschamps et al., 2023). Finally, extracting common species responses to environmental gradients from single SDMs is not straightforward since different models can have selected different variables. Joint species distribution models (jSDM) offer interesting avenues to address these issues (Pollock et al., 2014; Warton et al., 2015). By jointly modeling multiple species, jSDMs can borrow strength from more common or data-rich species that share similar environmental responses. This approach provides more robust parameter estimates for each species, including rare ones (Norberg et al., 2019). If many species show parallel responses to certain environmental gradients (e.g., precipitation), these gradient-response relationships can be learned more accurately by pooling data, and reveal ecological clusters or guilds—groups of species responding similarly to temperature, precipitation, or land cover. This reduces uncertainty in model estimates and can improve predictions for each individual species. Additionally, since jSDMs explicitly model the joint occurrence of species, they are supposed to yield more accurate estimations of community composition. Traditional jSDMs can be computationally cumbersome, especially those using purely Bayesian or likelihood-based approaches and latent variable implementations (LV-jSDMs) (Ovaskainen and Abrego, 2020). In addition, most jSDM are constrained by linear assumptions, requiring manual feature engineering to capture nonlinear relationships.

To overcome these limitations, we have designed the MTEC (Multi-Task modeling of Ecological Communities), a deep latent variable jSDM integrating multi-task neural networks to capture complex species-environment relationships (Hu et al., 2025) and variational autoencoders for scalable inference of latent factors (Tang et al., 2018). Using large-scale earthworm data from mainland France including Corsica, we compare the predictive performance of MTEC to that of traditional SDMs. We then identify groups of species with shared habitat preferences, by using the explainable AI framework of Shapley additive explanations SHAP (Lundberg and Lee, 2017) to attribute earthworm distribution patterns to specific environmental drivers. Combining joint species modeling with explainable AI enhances predictive accuracy while providing insights into the ecological processes shaping earthworm assemblages.

# 2. Materials and Methods

## 2.1. Overview

We modelled earthworm distributions using historical calibration datasets covering diverse environmental conditions and evaluated the models on contemporary datasets (Figure 1). The calibration dataset originated from Marcel Bouché's extensive surveys across mainland France including Corsica (Bouché, 1972), recently digitized and taxonomically updated by (Gérard et al., 2025). A separate dataset of recent georeferenced observations served as independent model evaluation data. Climate, soil, hydrology, and land cover variables were integrated as environmental predictors to capture major abiotic gradients. We implemented a deep joint species distribution model (MTEC), enabling simultaneous modeling of multiple earthworm species. Using the Shapley Additive Explanations (SHAP) framework, we identified and assessed key environmental drivers shaping species distributions. Lastly, unsupervised clustering was applied to categorize species into ecological response groups based on their environmental affinities.

## 2.2 Earthworm observation data

### 2.2.1 Calibration dataset

During the 1960s, Marcel Bouché collected extensive data on earthworm communities in mainland France (including Corsica), documented in detail along with habitat characteristics (geographical coordinates, vegetation, land use, soil measurements) in his seminal work (Bouché, 1972). The earthworm occurrence data (presence/absence) used here were digitized by (Gérard et al., 2025), resulting in 34,737 individual earthworm records from 1,399 sampling sites, covering 126 species and subspecies.

For modelling, we selected taxa occurring in at least five sites with detailed habitat descriptions. The refined dataset included 1,346 sites and 77 taxa (species and subspecies sensu Bouché, 1972).

### 2.2.2 Evaluation dataset

We aggregated a dataset of geotagged earthworm observations from multiple sources sampled in the [1998-2019] period. This dataset contains 1,126 observations of 37 out of the 77 taxa occurring in the calibration dataset. Due to the multiplicity of sources and the heterogeneous sampling effort, we treated this collection as presence-only data.

## 2.3 Environmental covariates

We assembled environmental predictors related to climate, soil chemical and structural properties, hydrology, and land cover type. Two distinct datasets were prepared for spatial and temporal alignment with earthworm occurrences for model training and projection:

The first dataset (1960-1970, calibration) integrated in-situ soil and land cover data from Bouché et al. (1972) and Gérard et al. (2025b), supplemented by GIS-derived environmental layers. Climate data were sourced from WorldClim (1960-1990; 1 km resolution), while soil and hydrological data were extracted from the European Soil Database at 1km resolution (see **Table 1** for more details about the specific variables we selected).

The second dataset (1990-2019, projection and evaluation) covering metropolitan France used climate variables from Worldclim, soil data from the European Soil Database, and land cover information from Corine Land Cover 2012-2018, maintaining the remaining predictors consistent.

## 2.4 Earthworm distribution models

### 2.4.1 Joint model of earthworm species

To model earthworm species distributions, we designed MTEC to simultaneously predict probabilities of species presence based on environmental covariates and latent factors representing unmeasured ecological processes (e.g., missing environmental variables, interspecies interactions). MTEC leverages neural network architectures to efficiently capture nonlinear species-environment relationships and improve multi-species prediction accuracy and computational efficiency simultaneously (**Figure 2**). See Supplementary Information for details on model formulation (SI, section A).

### 2.4.2 Training

We implemented MTEC in Python using Keras (Chollet, 2015). We trained MTEC (Figure 1a) using the Adam (Kingma and Ba, 2014) gradient-based optimization algorithm, minimizing a total loss composed of the species prediction loss (weighted binary cross-entropy, Goodfellow et al., 2016), the latent factor inference loss (Kullback-Leibler divergence, Kullback and Leibler, 1951) and a regularization term to control model complexity and prevent overfitting (Hastie et al., 2009; Hoerl and Kennard, 1970).

To improve the prediction accuracy of rare taxa, we used an imbalance-aware partitioning strategy (SI, section B2) to ensure their adequate representation in the training dataset. Model selection was based on $5 \times 2$ cross-validation (Dietterich, 1998), with Area Under the Receiver Operating Characteristic Curve (ROC-AUC) and the True Skill Statistics (TSS) as the primary evaluation metrics (SI, section B1).

### 2.4.3 Stacked species distribution models

For comparison with MTEC, single species distribution models based on Generalized Linear Models (Nelder and Wedderburn, 1972) , Gradient Boosting Machines (Elith et al., 2008), Random Forests (Breiman, 2001) were fitted independently using biomod2 in R (Thuiller et al., 2016, 2009) on the same calibration dataset as MTEC. Community composition predictions from these single-species models were combined through stacking (S-SDMs).

### 2.4.4 Projection and comparative evaluation

Using environmental data from the projection dataset, we predicted community composition across France at a 1 km resolution using MTEC (Figure 1b) and S-SDMs, producing habitat suitability maps for all species. Model extrapolation was evaluated by comparing predicted suitability to occurrence points in the evaluation dataset using recall scores.

Finally, we compared the predictive performances of the three S-SDMs to MTEC in terms of ROC-AUC and True Skill Statistics (TSS) on the calibration dataset and in terms of recall on the evaluation dataset using the Wilcoxon rank-sum tests (Wilcoxon, 1945).

## 2.5 MTEC interpretability

To identify the key environmental factors shaping earthworm distribution, we applied the Shapley Additive exPlanations (SHAP) framework (Lundberg and Lee, 2017) using the R package iml (Molnar, 2018). SHAP is model-agnostic and provides a detailed breakdown of how each environmental variable contributes to the predicted species probabilities in every single pixel of interest (Figure 1c). A positive SHAP value indicates that an environmental variable increases habitat suitability, while a negative value suggests a reduction compared to the average prediction. The effect of a variable can vary across its range—positive in some conditions, negative or neutral in others.

### 2.5.1 Attributing the environmental drivers of earthworm distribution

We computed the SHAP values for all environmental covariates for each species across samples of the calibration dataset (Figure 1c). From these, we generated various statistical summaries to derive ecological insights into earthworm distribution through the lenses of the model. The absolute value of the SHAP statistic reflects the *local importance* of an environmental factor at a specific site. Mapping these values geographically revealed spatial variability in influential factors. Averaging SHAP values yielded global measures of variable importance for each species (*global importance*).

### 2.5.2 Comparing earthworm species responses to the environment

Pairwise distances among species based on SHAP values were calculated, followed by hierarchical clustering (Ward, 1963). The optimal number of clusters was identified using the average of the GAP and WSS statistics (Tibshirani et al., 2001). Principal Component Analysis (PCA) was conducted on mean SHAP values per species, facilitating visualization and interpretation of species response groups. We particularly focused on precipitation and land cover, given their known significance in structuring earthworm communities and relevance under global environmental changes (Phillips et al., 2019).

## 3. Results

### 3.1. Predictive performances

Comparison between MTEC and single-species distribution models revealed significantly improved predictive performance of MTEC on the calibration dataset (Table 2). Predictive performance was generally lower for widespread taxa and variable for less prevalent taxa (SI, section C). Specifically, the lowest ROC-AUC values (~0.69) of MTEC were obtained for widely distributed taxa such as *Lumbricus castaneus disjonctus* Tétry, 1936 and *Octolasion cyaneum* (Savigny, 1826).

For highly prevalent species such as *Aporrectodea caliginosa (Savigny, 1826)* and *Lumbricus terrestris* Linnaeus*, 1758,* performance differences between MTEC and more advanced SDMs (RF and GBM) were marginal or slightly unfavorable to MTEC.

This indicates that common species benefit less from shared environmental information than rare or less common taxa. However, MTEC demonstrated a higher ability (+0.4) to predict contemporary occurrences from the evaluation dataset (Table 2), suggesting that single-species SDMs may be more prone to overfitting.

### 3.2. Habitat suitability

#### 3.2.1. Abiotic drivers of earthworm habitat suitability

Land cover, precipitation extremes, and temperature seasonality emerged as the most influential abiotic factors across taxa (Figure 3). Interestingly, total annual precipitation had limited predictive value, whereas specific precipitation metrics such as rainfall during the wettest month (typically autumn in France), and precipitation during the coldest (winter) and warmest (summer) quarters were critical, along with precipitation seasonality. Among soil properties, soil pH was ranked among the most important variables, while organic matter quality (C/N ratio) influenced suitability more than the quantity of soil organic carbon. The soil water regime strongly affected habitat suitability for many taxa, whereas proximity to water bodies was largely insignificant, likely due to the coarse scale of hydrological rasters. Additionally, soil bulk density had a greater importance than soil texture.

The contribution of environmental variables can be assessed spatially. As an illustration, Figure 4 shows the geographic distribution of the SHAP values, which can be interpreted as the local partial effect, of the most important climate variables for *Allolobophora icterica*. (See SI, section C3 for other modeled taxa).

### 3.2.2 Response groups of earthworms to environmental gradients

Focusing on precipitation and land cover, two major ecological drivers highlighted in previous studies, we visualized species-specific responses through multi-species SHAP summary plots.

The PCA on SHAP values related to precipitation explained 60% of the total variation (Figure 5). Axis 1 (43% of variance) differentiated species based on their affinity to climates with varying precipitation seasonality, with high seasonality associated with negative PCA scores. Using hierarchical clustering informed by the GAP statistic, we identified four response groups:

- The first cluster predominantly included Mediterranean species, including *Scherotheca porothec*a (Bouché 1972), *Hormogaster samnitica* (Cognetti, 1914), *Kritodrilus calarensis* (Tetry, 1944), or species within the genus *Diporodrilus*.
- The second cluster comprised numerous endemics of the South of France and the Massif Central, from genera *Zophoscolex*, *Prosellodrilus*, *Scherotheca*, *Gatesona* and *Hemigastrodrilus*, along with eurytopic species such as *Allolobophora chlorotica* (Savigny, 1826) or *Aporrectodea rosea* (Savigny, 1826).
- The third cluster included taxa like *Eiseniella tetraedra* (Savigny, 1826), *Satchellius mammalis* (Savigny, 1826), *Dendrobaena octaedra* (Savigny, 1826), *Aporrectodea giardi* (Ribaucourt, 1901), and *Lumbricus centralis* Bouché, 1972.
- The fourth cluster consisted primarily of widespread species (e.g., *Ap. caliginosa*, *O. cyaneum*) and a species known for riparian preferences (*Haplotaxis gordioides* (Hartman, 1821)).

Regarding land cover, the PCA on SHAP values accounted for approximately 80% of the overall variation (Figure 6). Axis 1 (62% of the variance) distinguished species preferring closed canopy habitats (negative scores) from those favoring open habitats (positive values) whereas axis 2 (17.5% of the variance) relates to species affinity to soil wetness. Clustering identified six distinct ecological response groups, corresponding to specialists of forests (cluster 6), shrublands (cluster 3), wetlands (cluster 5) and three grassland clusters: meadows (cluster 1), wet grasslands (cluster 2) and pastures (cluster 4). Taxa were unevenly distributed among these groups. Shrubland and grassland specialists each comprised 15–20 taxa, while forest and pasture specialists were less numerous (~10 taxa). Only four taxa were classified as wetland specialists. However, some clusters presented affinities with different habitats (e.g. cluster 2).

A similar analysis was conducted for earthworm response groups to temperature and soil properties (SI, section C4).

# 4. Discussion

In this study, we advance earthworm species distribution modeling by jointly predicting the distributions of 77 taxa across France, revealing more precise environmental drivers of their distribution. Using MTEC, an innovative joint Species Distribution Model (jSDM) powered by neural networks, we confirmed that multi-task neural networks notably outperform traditional single-species SDMs, especially in modelling rare taxa (SI, section C1). Additionally, our analysis identified primary environmental drivers shaping earthworm taxa distributions and used clustering to illustrate distinct ecological response groups.

### A deep joint species distribution model

MTEC represents an evolution in jSDM techniques, combining neural networks with latent variables in a probabilistic framework akin to generalized linear latent variable models (GLLVM) (see Knape (2016) and Harris (2015)). Unlike conventional jSDMs employing generalized linear models, MTEC leverages neural networks that combine the functional representation capabilities of neural networks with the inferential capacity of hierarchical models to accommodate complex features and model non-linear species-environment relationships. The use of variational autoencoders (VAEs) within MTEC facilitates efficient latent variable modeling with fixed complexity relative to dataset size. Although VAEs inherently assume independent latent factors, recent developments propose spatially and temporally structured latent priors (e.g., Gaussian Process Priors, Kingma and Welling (2022)), potentially enhancing ecological modeling applications. Beyond undirected association networks, VAEs have also been applied in domains outside ecological research to learn the structures of Bayesian Networks (Zhang et al., 2019) and Gaussian Copulas (Mazoure, 2019).

**Predictive performances**

Contrary to previous reviews indicating limited advantages of jSDMs over stacked SDMs (Pichler and Hartig, 2021; Wilkinson et al., 2019), our findings demonstrate that MTEC frequently surpasses single-species models. MTEC's superior predictive performance, particularly for rare taxa, is likely due to shared environmental feature extraction, increased statistical power, and targeted loss weighting strategies addressing data imbalance. However, a few widespread taxa showed relatively low predictive performance across all fitted methods, suggesting that forecasting habitat suitability is more challenging for species with high adaptive capacities. This could be due to missing relevant environmental drivers (e.g. soil disturbance). For such widespread species which also occur beyond our study extent (France), it is likely that a larger scale modeling is required to cover their full niche distribution. Modeling relative abundances, rather than occurrences, could provide a deeper understanding of these species' ecological niches. In some instances, poor model performance might also indicate the presence of cryptic species creating taxonomic confusion in available data (e.g. among *L. terrestris* and *Lumbricus herculeus* (Savigny, 1826; James et al., 2010), which further complicates accurate predictions (Marchán et al., 2020). Integrative taxonomy, which combines morphological, molecular, ecological, and geographical data, offers a promising approach for resolving complex species boundaries, especially in soil biodiversity. Coupled with AI, it can accelerate species identification and pattern discovery across large, multidimensional datasets (Karbstein et al., 2024).

Beyond improving taxonomic discrimination, future research should integrate finer-scale environmental data that describe microhabitats to improve soil biodiversity modeling. Examples include proximal sensing data like soil electrical conductivity (EC) and near-infrared absorbance (NIR) (Schirrmann et al., 2016). Proxies for vegetation structure and temporal variability, such as NDVI-based metrics, along with soil depth and other topographic features like slope, topographic wetness index, and solar radiation, could be obtained via airborne or satellite remote sensing (Ou et al., 2021).

### Abiotic drivers of habitat suitability for earthworm species

The relative importance of environmental factors and the nature of species' responses differed among taxa. Consistently with prior research, land cover, precipitation, and temperature emerged as key determinants of earthworm habitat suitability (Phillips et al., 2019; Rutgers et al., 2016). Although our study operates at a mesoscopic scale (France) compared to Phillips et al. (2019), the geographic breadth of our work encompasses diverse climates and land uses, allowing for a clear hierarchy to emerge: at fine scales (landscape, catchment), soil properties dominate, whereas at broader scales, where climate and land use shift significantly, climate becomes the primary driver of species distributions.

Precipitation regimes influence soil moisture, which in turn affects the activity and development of earthworms. Eisenhauer et al. (2014) demonstrated through experimental warming that earthworm development is constrained by temperature only when it leads to drought conditions. Rather than total precipitation, specific seasonal metrics (e.g., wettest month precipitation) proved crucial. While extreme precipitation events are known to affect earthworms (Singh et al., 2019), the significance of seasonality has led some species to develop strategies for coping with high climate variability both during the day and across months. Conversely, several species, primarily epigeic ones, clearly prefer more uniform and optimal climatic conditions, such as observed in regions like northwestern France.

At the edaphic level, soil pH was ranked among the most important variables, aligning with previous knowledge (Bouché, 1972). Soil pH influences not only plant growth and litter production but also the availability of metals (Al, Zn, Cu, Cd) that can reach toxic levels for earthworms (Spurgeon and Hopkin, 1996). The soil water regime was highly significant for most taxa, while proximity to water courses had a negligible impact, either due to the coarse scale of the raster layers, or its redundancy with soil water regime and wetland variables. Additionally, soil bulk density had a greater effect at the assemblage level compared to texture, a result that is not surprising given that bulk density was estimated from soil texture and structure classes using pedo-transfer rules (Hiederer, 2013). This transformation may have resulted in some loss of information and could have reduced the discriminative power of the texture features. Nevertheless, texture remained the most distinguishing factor for species such as *Dendrodrilus subrubicundus* (Eisen, 1874) and *E. tetraedra*, which are dependent on sandy soils near water bodies.

**Building groups of response to environmental gradients**

In the context of European efforts to monitor and safeguard soil life (e.g., EU soil strategy for 2030), there is growing recognition of the need for robust, interpretable metrics that can inform management and policy decisions (Eisenhauer et al., 2022). One promising approach is to classify taxa into ecological response groups, such as habitat specialists vs. generalists or disturbance-sensitive vs. tolerant species. This strategy, already highlighted by Gérard et al. (2025a) in the development of new biodiversity indicators, can help detect shifts in community composition that are not visible through richness-based metrics alone. Similarly, Jupke et al. (2024) have emphasized the relevance of identifying typical earthworm assemblages across land-use gradients, offering a more integrative view of how soil fauna respond to environmental change. By building such response groups, we aimed to move beyond taxonomic inventories and towards ecological interpretations that are both actionable and scalable. We illustrated this approach with response groups to precipitation and land use.

Regarding precipitation regimes, amongst the four identified species groups, the first cluster primarily consisted of Mediterranean species that exhibit a strong ability to thrive in warm ecosystems with highly seasonal rainfall, such as *H. samnitica* and *K. calarensis*. Species within this cluster demonstrated a positive response to precipitation seasonality, likely avoiding adverse climate conditions by their ability to enter diapause in the deeper layers of the soil (Bouché, 1972). The second cluster included many species from the genera *Zophoscolex*, *Prosellodrilus*, *Scherotheca*, *Gatesona* and *Hemigastrodrilus*, which are endemic to Southwestern Europe. It also comprised some eurytopic, widespread species like *All. chlorotica* and *Ap. rosea*. We interpreted this cluster as a group of mesophilous, warmth-tolerant species. The third cluster included species such as *E. tetraedra*, *S. mammalis*, *D. octaedra*, *Ap. giardi*, and *L. centralis*. These species require moist conditions for most of the year to thrive (Bouché, 1972). The final cluster consisted of a few species, some of which are known to be riparian. These species share a high affinity for a precipitation regime that is uniformly distributed during the year. Climate models project an increase in the frequency of extreme precipitation events, while the total amount of rainfall is expected to remain comparable to its current state, with greater variability. Given earthworm sensitivity to changes in precipitation seasonality, these alterations are likely to impact their distributions. Rising temperatures, drought, and increased winter rainfall will influence soil moisture and temperature regimes, each affecting earthworms differently. The combination of elevated temperatures, which heighten metabolic demands, and increased extreme precipitation events may alter earthworm life cycles. Additionally, extreme rainfall and changes in land use can make soils more vulnerable to erosion, destroying earthworm habitats (Singh et al., 2019).

The classification procedure on land cover response revealed significant differences in earthworm habitat preferences. We identified six response groups, five of which corresponded to open ecosystems and three of which were associated with very distinct land cover classes: forests, pastures, and inland wetlands. Taxa with affinities for shrub and herbaceous vegetation were divided into two groups along the forest-to-pasture gradient: scrub woodlands and meadows. An additional intermediate group was identified for species from wet grasslands. Most taxa exhibited a neutral or negative response to forest soils, with the exception of a limited number of litter-dwelling species, such as *D. octaedra*. A few riparian species, such as *E. tetraedra* and *H. gordioides,* and several endogeics, e.g. *O. cyaneum* or *Helodrilus oculatus* (Hoffmeister, 1845), demonstrated a strong preference for inland wetlands. Other litter-dwelling species were positively associated with pastures, which may support a higher species richness (Decaëns et al., 2008). Shrubs and meadows were co-dominated by endogeic and anecic worms. Forests, pastures, and shrub/herbaceous land cover classes accounted for approximately 80% of the calibration dataset. In contrast, arable and permanent crop lands were underrepresented in the training dataset. Their composition included widespread species. The lack of epigeics is currently reported to be linked to farming practices such as tillage that impact surface soil fauna (Briones and Schmidt, 2017; Decaëns et al., 2008). There were no species with a strong affinity for agricultural sites, although a few demonstrated a strong aversion toward these systems (negative SHAP values). These conclusions may vary when considering relative abundances rather than presence/absence data.

Both climate conditions and land cover emerged as key predictors in our model, underscoring their combined influence on earthworm distributions. Anticipating future shifts will therefore require the development of scenarios that integrate the interplay between changing precipitation regimes and land cover dynamics. Beyond this, we demonstrated the potential of classifying species into response groups along two major environmental gradients, offering a framework to interpret biodiversity patterns more effectively. Such classifications could be extended to encompass other critical constraints influencing earthworm populations, providing further ecological insight.

## Conclusions and future directions

Understanding the drivers of earthworm distribution is essential for predicting their response to ongoing global changes and estimating the spatial emergence of their functional benefits as ecosystem engineers and trophic resources for wildlife. So far, most studies have utilized single species distribution models that focus on a limited number of widespread taxa or, conversely, aggregated indicators such as biomass or richness, without considering species identities. Studying each species separately using dedicated species distribution models (SDMs) can be time-consuming and may be impractical for species with few available observations due to their intrinsic rarity or an insufficient sampling effort. The multi-task model presents new opportunities to study earthworm responses to global change. Future improvements in earthworm distribution modeling should incorporate biotic interactions and microhabitat characteristics, alongside advanced remote sensing methods, to better capture complex ecological patterns (Hu et al., 2025; Poggiato et al., 2025). Integrating species dispersal capabilities through spatially explicit modeling will further enhance the accuracy of predicted distributions under global environmental changes. Ultimately, multi-species distribution models like MTEC represent valuable tools for understanding earthworm biodiversity dynamics and guiding effective conservation and management policies in soil ecosystems.

## Data and code availability statement

All data supporting the findings of this study are available within the article and its supplementary materials. Code used to run the analysis is available through a GitHub repository: https://github.com/bettasimousss/DeepLombrics.git

## Author contributions

MH designed the study. SS-M developed the analytical framework with guidance from EG and WT. MH, TD, YC, and SG provided the earthworm datasets. SL and colleagues contributed the environmental data layers. TD, MH, DM, YC, SG and CM participated in the interpretation and ecological contextualization of the results. MH, and SS-M drafted the initial manuscript. All authors contributed to manuscript revisions and approved the final version.

## Funding statement

This research is a by-product of the IMPACTS and LandWorm groups funded by the Centre for the Synthesis and Analysis of Biodiversity (CESAB) of the Foundation for Research on Biodiversity (FRB) and the Ministry of Ecological Transition. WT and SS-M. also acknowledge support from the HorizonEurope OBSGESSION (N°101134954) and NaturaConnect (N°101060429) projects and the MIAI@Grenoble Alpes (ANR-- 19-- P3IA-- 0003). SG PhD thesis was supported by the ENS and by the INRAE Agroecosystem division through the GloWorm project.

# Conflict of interest

The authors declare no conflict of interest.

# Ethics approval statement

This study did not involve human participants or experimental animals and therefore did not require ethics approval.

# Tables

**Table 1:** List of environmental predictors, definition, units and sources (calibration and projection).

| Group | Variable | Description | Unit, Type | Source (calibration) | Source (projection) |
|---|---|---|---|---|---|
| Temperature | TMeanY | Annual mean temperature | Celsius | WorldClim (1960-1990) (Fick and Hijmans, 2017) | CHELSA (1979-2013) (Karger et al., 2017) |
| | TMeanRngD | Mean diurnal range | | | |
| | Tiso | Isothermality | | | |
| | Tseason | Temperature seasonality | | | |
| | TMaxWarmM | Maximum of Warmest Month | | | |
| | TMinColdM | Minimum of Coldest Month | | | |
| | TRngY | Annual range | | | |
| | TMeanWetQ TMeanDryQ TMeanWarmQ TMeanColdQ | Mean temperature of Wettest/Driest/Warmest/Coldest Quarters | | | |
| | TMinColdM TMaxWarmM | Min/Max temperature of Coldest/Warmest Months | | | |
| Precipitation | PTotY | Annual precipitation | mm | | |
| | PSeason | Precipitation seasonality | | | |
| | PWetM PDryM | Precipitation of Wettest/Driest Month | | | |
| | PWetQ PDryQ PWarmQ PColdQ | Precipitation of Wettest/Driest/Warmest/Coldest Quarters | | | |
| Land cover | LandCover | Land cover class | Categorical (16 levels) | (Bouché, 1972) | Corine land cover (EEA 2018) |
| Soil chemical properties | pH | Water acidity level (pH) | None | (Bouché, 1972) | ISRIC soilgrids (Batjes et al., 2020) |
| | C | Organic carbon (C) | mg/Kg soil | | |
| | C/N | Carbon to Nitrogen ratio (C/N) | No unit | | |
| Hydrological properties | ProxiWater | Proximity to watercourses of at least 50 m width | Binary | (Botella et al., 2019) | (Botella et al., 2019) |
| | AWC | Available water capacity of topsoil | mm | European soil database v2 raster library (ESDB) (Panagos et al., 2012) | European soil database v2 raster library (ESDB) (Panagos et al., 2012) |
| | BaseSat | Base saturation of topsoil | % | | |
| | CEC | Cation exchange of topsoil | cmol/Kg soil | | |
| | WaterRegime | Water regime | Categorical - 5 levels | | |
| Soil structural properties | Clay | Soil clay content | % | (Bouch´é, 1972) | ISRIC Soilgrids rasters (Batjes et al., 2020) |
| | Silt | Soil silt content | % | | |
| | Sand | Soil sand content | % | | |
| | Crusting | Soil crusting class | Categorical (6 levels) | ESDB (Panagos et al., 2012) | ESDB (Panagos et al., 2012) |
| | Erodi | Erodibility class | Categorical (6 levels) | | |
| | DepthGleyH | Depth to gley horizon | cm | | |
| | DepthImpL | Depth to impermeable layer | cm | | |
| | ParentMat | Dominant parent material | Categorical (9 levels) | | |
| | SBD | Soil bulk density | g/cm3 | | |

**Table 2:** Median and standard deviation of the predictive performance metrics on the cross-validated calibration dataset (TSS, ROC-AUC) and on the independent evaluation dataset (Recall) across the 77 modeled taxa, comparing the proposed multi-task architecture (MTEC) with single species distribution models (GBM, GLM, and RF).

| Model | TSS | ROC AUC | Recall (Evaluation) |
|---|---|---|---|
| MTEC | 0.746 ± 0.243 | 0.912 ± 0.102 | 0.914 ± 0.214 |
| GLM | 0.411 ± 0.234 | 0.681 ± 0.186 | 0.519 ± 0.389 |
| GBM | 0.552 ± 0.281 | 0.753 ± 0.238 | 0.302 ± 0.267 |
| RF | 0.646 ± 0.247 | 0.816 ± 0.145 | 0.250 ± 0.314 |

# Figures

**Figure 1 - Overview of the earthworm species distribution modeling study.**

(a) Training: a deep-learning joint species distribution model (MTEC) is trained on earthworm communities sampled across France. The model integrates in-situ field descriptions (land cover, soil physico-chemical properties) with environmental GIS databases (climate, hydrography, pedology). These inputs are embedded by a neural network into a shared environmental representation, along with latent factors capturing missing predictors, unmeasured features and interspecific associations to optimize multi-species predictions.

(b) Projection and evaluation: The trained model is applied to continuous environmental layers to predict habitat suitability for earthworm species across France. The resulting habitat suitability maps are evaluated using independent observations to assess the accuracy of the predictions.

(c) Explainability: Model predictions are analyzed using SHAP (SHapley Additive exPlanations). For a given set of observations (calibration dataset), SHAP values quantify the positive or negative impact of each environmental feature on the habitat suitability of each species within each site. Statistical summaries of these SHAP values help identify local and global drivers (SHAP importance), habitat preferences (SHAP summaries), and response groups (SHAP clustering).

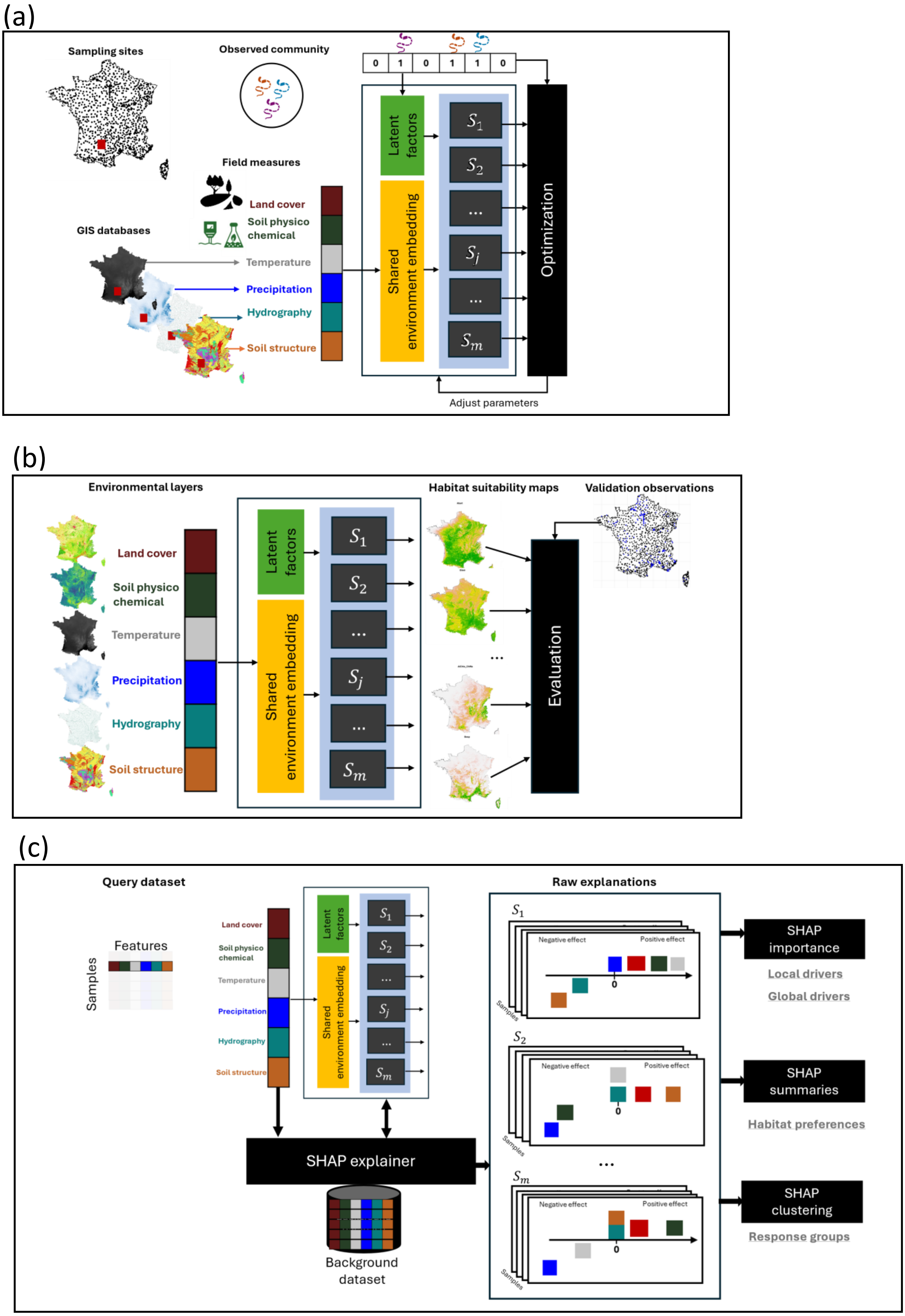

**Figure 2: Architecture of MTEC**. MTEC consists of three main modules: a probabilistic encoder, an environmental feature encoder, and a multi-task decoder. The probabilistic encoder is a fully connected neural network that learns a variational posterior distribution $q$ over latent factors $h$, capturing unobserved ecological influences. The environmental feature encoder processes raw environmental covariates $e$ through a separate fully connected network, extracting a shared embedding $x$ relevant to all species. This introduces non-linearity in the response and reduces the need for feature engineering. Finally, the multi-task decoder predicts species occurrence probabilities $\theta$ as a function of the shared environmental embedding (response coefficients $\beta$) and the latent factors (latent factor loadings $\alpha$). During training (a), latent factors are inferred from the approximate posterior distribution $q$ given known community data $y$. During prediction (b) on new sites, the latent factors are sampled from the prior distribution $p$.

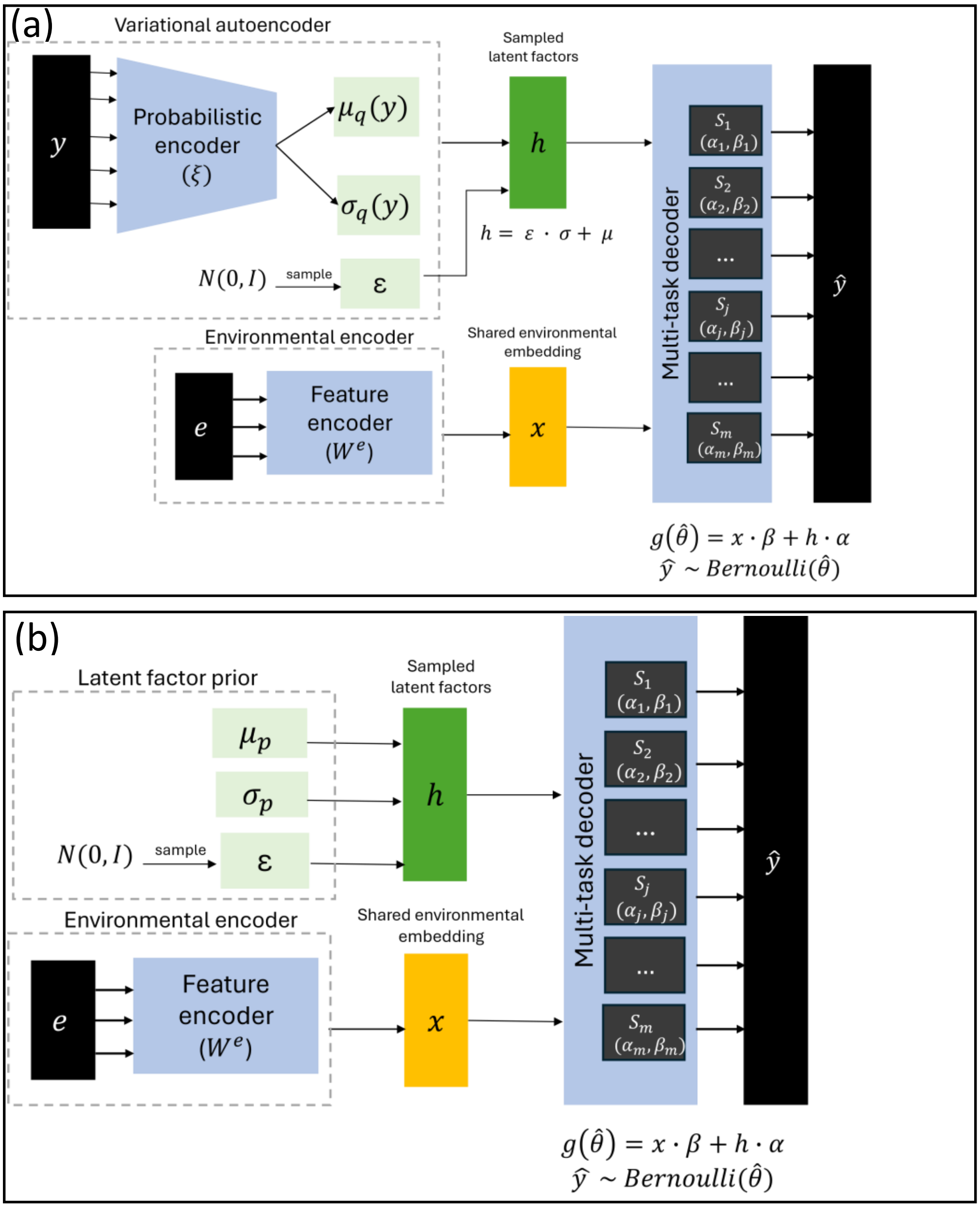

**Figure 3: Global environmental feature contributions.** For each taxon, the importance of each environmental factor was quantified as the average (SHAP) contribution of that variable to habitat suitability probability across all sites. (a) The distribution of environmental features importance among taxa is presented in descending order, highlighting the relative influence of each variable on the model's predictions. Meaning of the codes for environmental features are given in Table 1. (b) Individual feature contributions were aggregated at the feature group level and normalized to obtain a variation partitioning for each earthworm species.

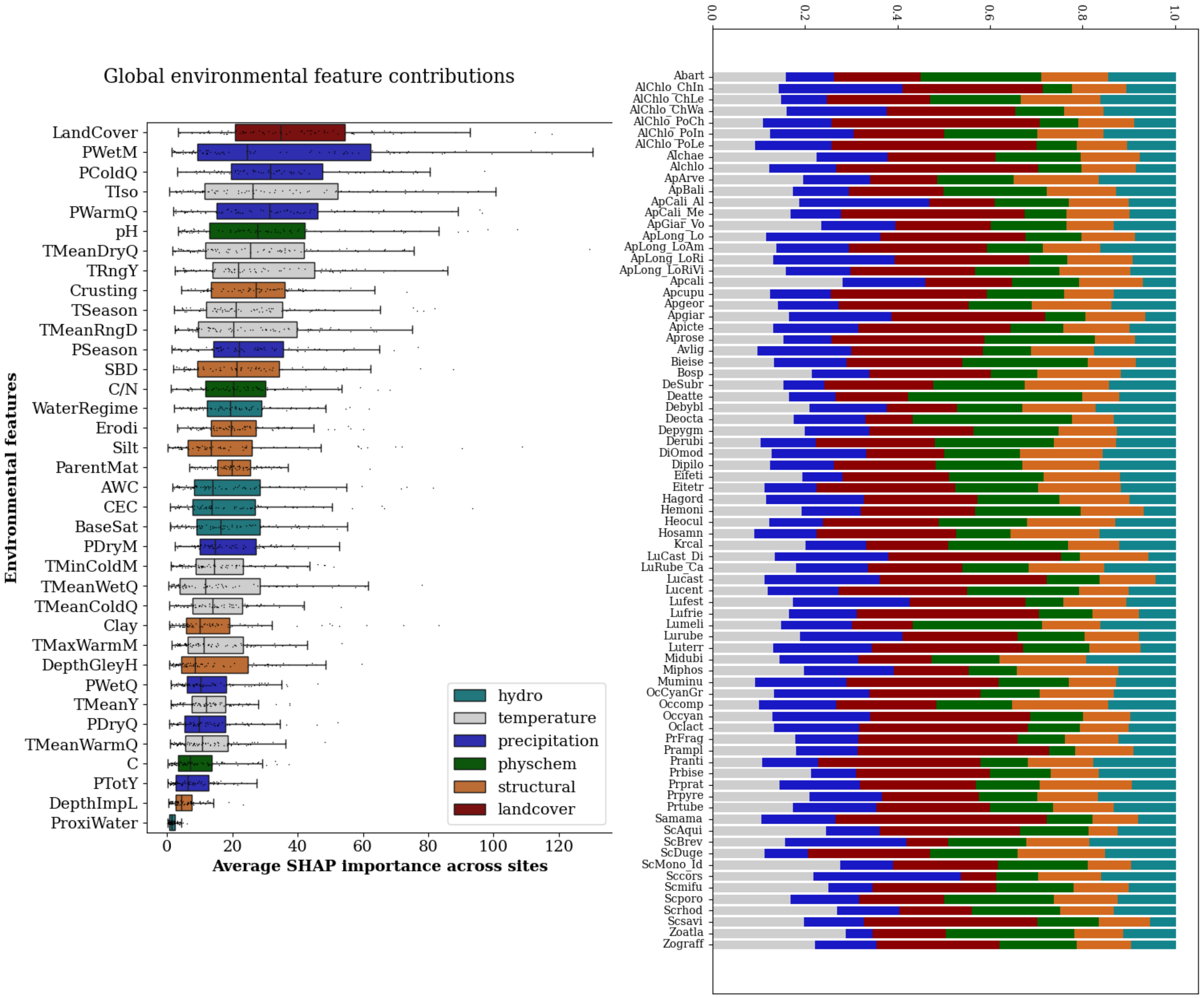

**Figure 4. Geolocalized contribution of environmental factors to explaining the occurrence of *Aporrectodea icterica* (Savigny, 1826).** This species is mostly influenced by precipitation and temperature factors, then structural, physico-chemical factors and land cover. For instance, in the mediterranean region, *A. icterica* presence is limited by the mean diurnal temperature range (TMeanRngD), the minimum temperature of the coldest month (TMinColdM), the precipitation of the warmest Quarter (PWarmQ) and the precipitation of the driest month (PDryM) and positively influenced by the precipitation during the coldest quarter (PColdQ). On the Channel facade, the factors with most positive importance were the mean temperature of the warmest quarter (TMeanWarmQ), the precipitation of the coldest Quarter (PColdQ) while those with negative influence were the isothermality (TIso), the mean daily temperature range (TMeanRngD) and the temperature seasonality (TSeason).

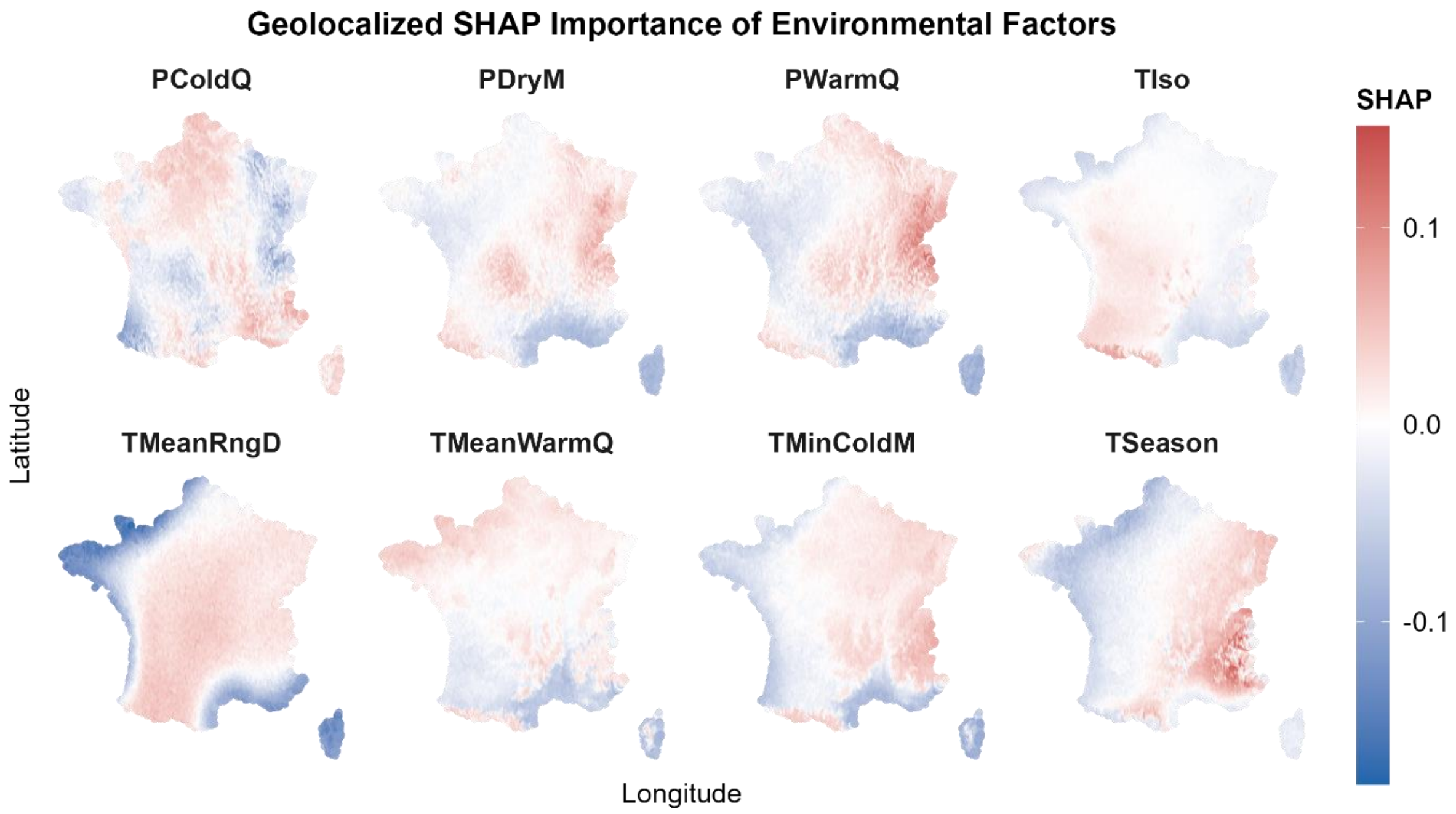

**Figure 5: Response of earthworms to precipitation. (a) Response groups to precipitation**: SHAP contributions to precipitation covariates across all observations (sites) for all species were organized into a matrix (species × site-covariate). Hierarchical clustering (Ward's algorithm) was applied to group species with similar responses across sites and precipitation covariates, with the GAP statistic identifying four optimal clusters. To aid visualization, Principal Component Analysis (PCA) was applied to reduce the dimensionality of the input matrix. The projection of species onto the first two principal components (explaining 56.9% of the variation) is shown, along with their assigned clusters. **(b) Multi-species SHAP summary plot of Precipitation Seasonality (PSeason)**: This plot visualizes how Precipitation Seasonality influences the habitat suitability of different earthworm taxa. Species label colors correspond to cluster colors (panel a). The y-axis shows the SHAP contribution, indicating whether PSeason has a positive or negative effect on predictions. A color gradient represents the range of PSeason values, helping to identify species preferred / limiting PSeason levels.

(a

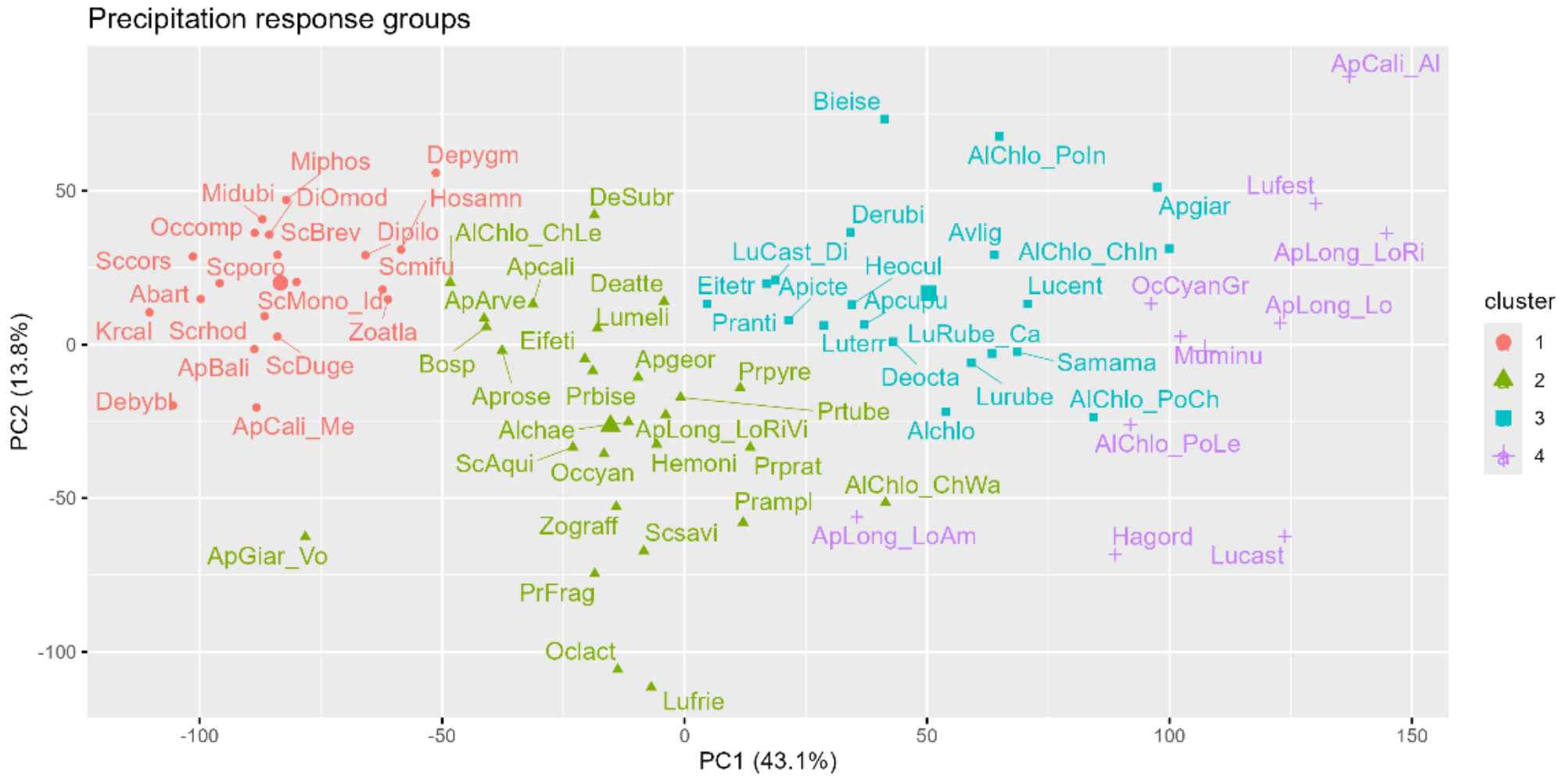

(b

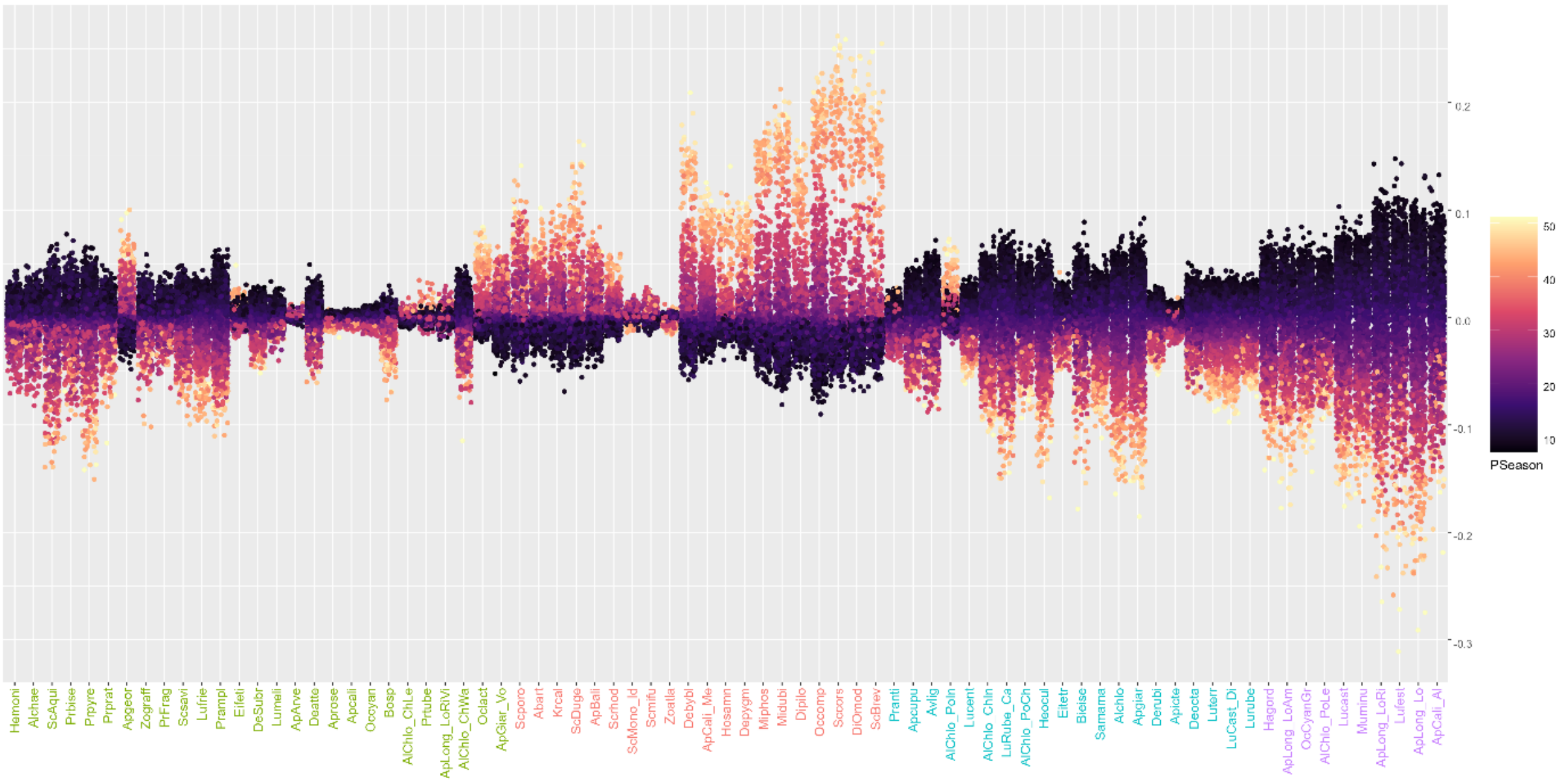

**Figure 6: Response of earthworms to land cover.** (a) **Earthworm response Groups to Land Cover** – SHAP contributions across sites were structured into a species × site matrix, with hierarchical clustering identifying six response groups (GAP statistic). To aid visualization, PCA was applied, and the projection of species onto the first two principal components (explaining 79.5% of the variation) is shown along with their assigned clusters. (b) **Multi-Species SHAP Summary Plot** – This plot visualizes the effect of land cover categories on different earthworm species. The y-axis represents SHAP contributions, indicating whether a land cover category (scatter point colors) has a positive or negative effect on species predictions helping to identify species preferential habitats. Species label colors correspond to the cluster colors (panel a).

(a)

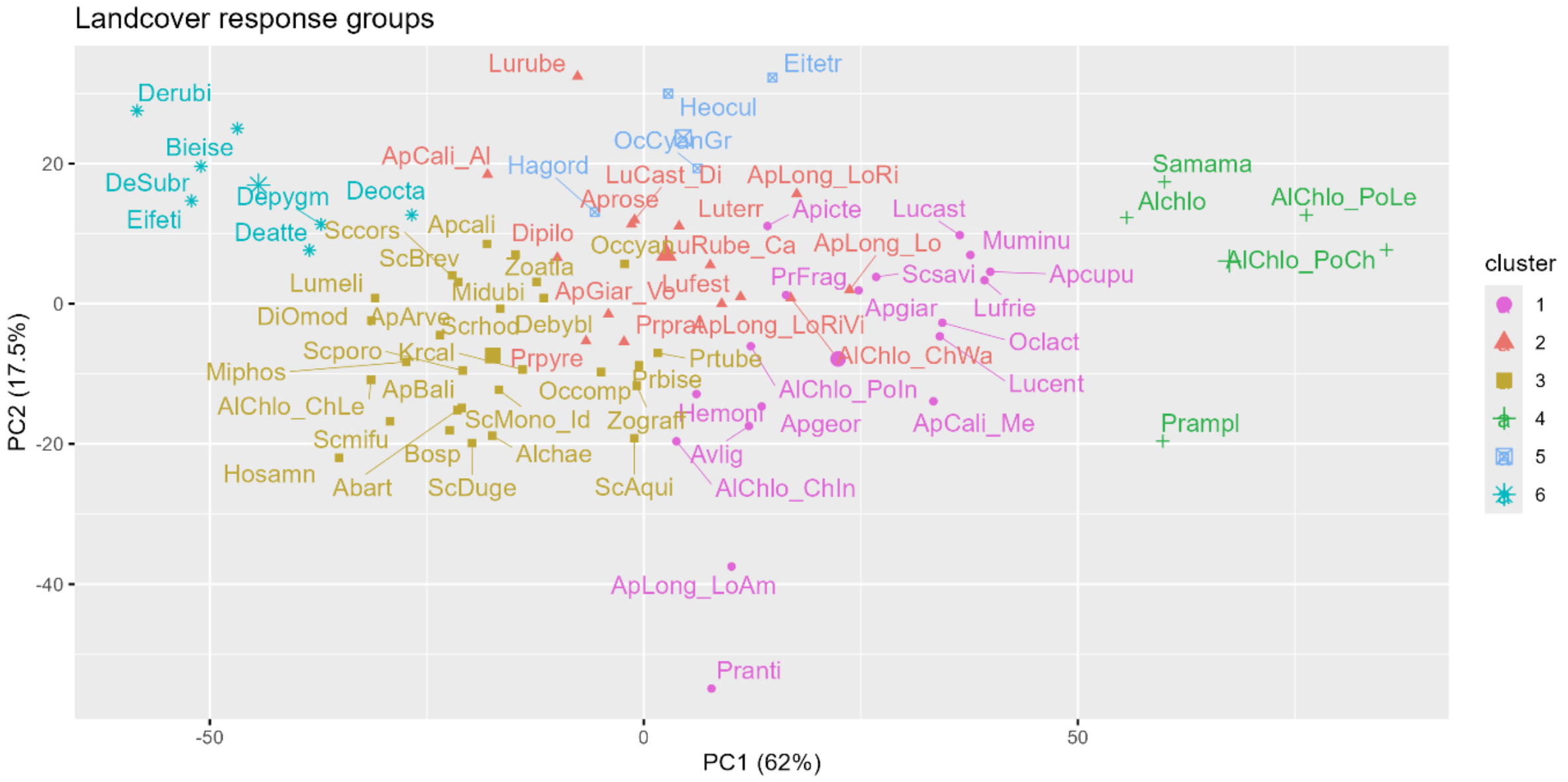

(b)

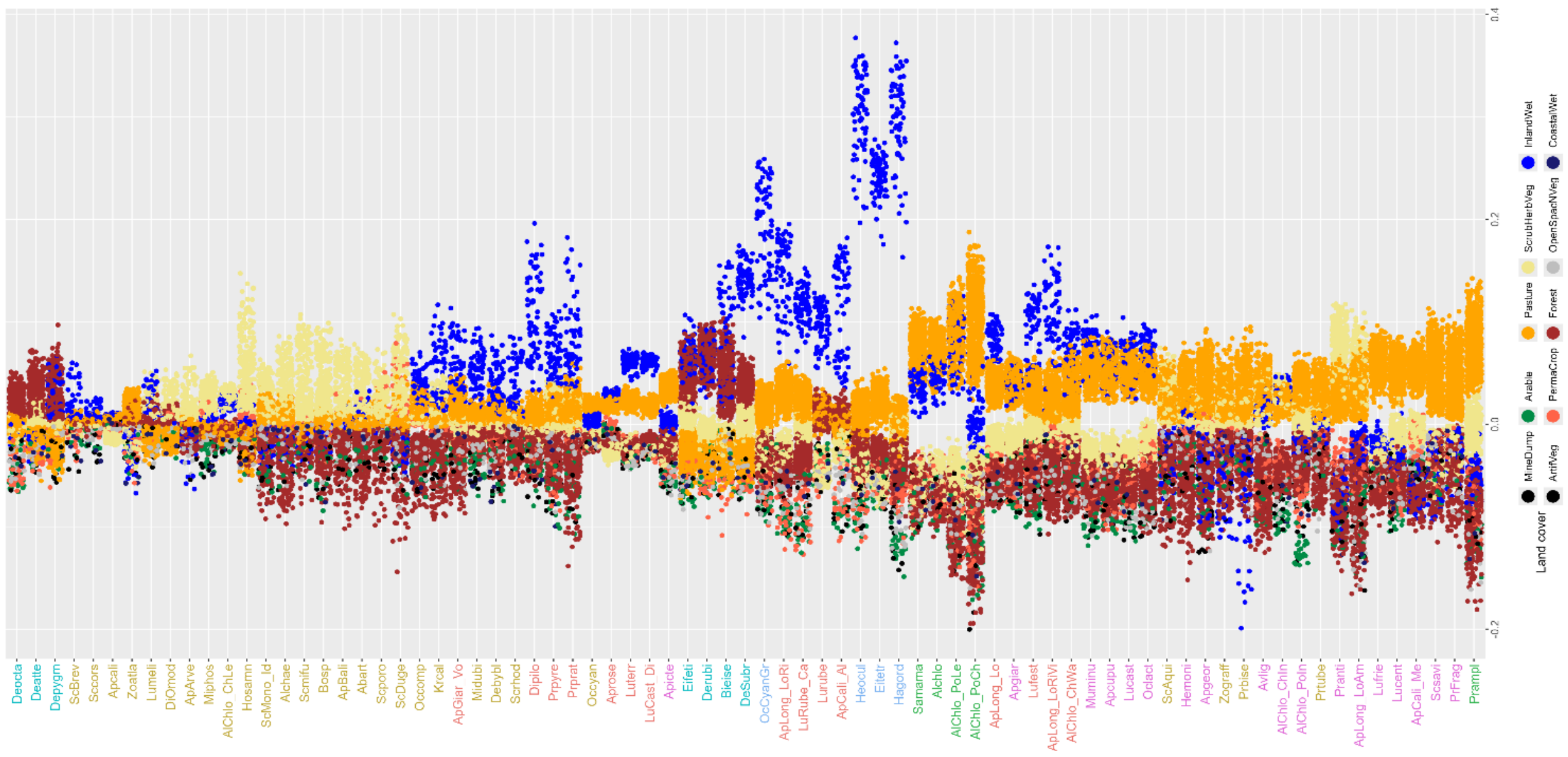

# Supplementary Materials

# Digging deeper: deep joint species distribution modeling reveals environmental drivers of Earthworm Communities

# A. Multi-Task modeling of Ecological Communities (MTEC)

Multi-Task modeling of Ecological Communities (MTEC) is a latent variable joint species distribution model that leverages neural networks as both feature extractors and inference tools. MTEC combines multi-task (MT) neural networks for flexible multi-species environmental responses and variational autoencoders (VAE) for scalable latent factor estimation.

The goal of MTEC is to predict the probability of occurrence of a set of target species at a certain location given observed environmental covariates and unobserved latent variables.

In section 1.1, we describe the mathematical details of MTEC. In section 1.2, we describe the neural network architecture used to fit the model to environmental and community data.

## A1. Mathematical details

### A1.1 Notation table

*Observed data*

This represents variables that are observed in the dataset.

| Symbol | Description |
|---|---|
| $\mathbf{i} \in \{\mathbf{1}, \ldots, \mathbf{N}\}$ | Training sites, where $N$ is the number of sites. |
| $\mathbf{j} \in \{\mathbf{1}, \ldots, \mathbf{M}\}$ | Target species, where $M$ is the number of species considered. |
| $\mathbf{E} = (\mathbf{e_i})_{\mathbf{i},}$ | Raw environmental predictors matrix, where $e_i$ is the set of raw environmental covariates at the $i^{th}$ site. |
| $\mathbf{Y} = (\mathbf{y_{ij}})_{\mathbf{i,j}}$ | Earthworm community matrix, where $y_{ij}$ is a binary indicator of species $j$ presence at site $i$. |

*Table 1 - Table of observed data symbols*

*Hidden variables*

This represents internal variables computed by the model such as feature embeddings or latent factors.

| Symbol | Description |
|---|---|
| $\mathbf{l} \in \{\mathbf{1}, \dots, \mathbf{L}\}$ | latent factors, where $L$ is the dimension of the latent space. |
| $\boldsymbol{H} = (\boldsymbol{h}_{il})_{i,l}$ | Hidden factors matrix, where $h_{il}$ is the value of the $l^{th}$ factor at site $i$. |
| $\boldsymbol{X} = (\boldsymbol{x}_{ik})_{i,k}$ / $\mathbf{k} \in \{\mathbf{1}, \dots, \mathbf{K}\}$ | Shared environmental embedding vector, where $K$ is the dimension of the embedding vector. |

*Table 2 - Table of hidden data symbols*

*Model parameters*

This represents parameters estimated by the model given data.

| Symbol | Description |
|---|---|
| $\mathbf{B} = (\boldsymbol{\beta}_{\mathbf{k,j}})_{\mathbf{k,j}}$ | Coefficient of species response to the (shared representation) environment |
| $\mathbf{A} = (\boldsymbol{\alpha}_{l,\mathbf{j}})_{l,\mathbf{j}}$ | Latent factor loading matrix |
| $\boldsymbol{W^e}$ | Parameters (weights, biases) of the feature extractor neural network |
| $\boldsymbol{\xi}$ | Parameters (weights, biases) of the probabilistic encoder (recognition) network |

*Table 3 - Table of model parameters symbols*

*Model settings*

These settings are fixed before the model begins training.

| Symbol | Description |
|---|---|
| $\boldsymbol{\mu_p}$ | Prior mean of the latent factors |
| $\boldsymbol{\sigma^2}_{\boldsymbol{p}}$ | Prior standard deviations of the latent factors |
| $\boldsymbol{g}$ | Link function for the species response |
| $\boldsymbol{L}$ | Dimension of the latent space |

*Table 4 - Table of model hyperparameters symbols*

## A1.2 Generative model

A deterministic neural network (feature extractor) $f^{ext}$ transforms raw input covariates into a more informative vector representation (eq. 1), shared across species and referred to as environmental embedding, which captures relevant features driving the community composition while reducing noise and irrelevant variability in the raw input.

$$x_i = f^{ext}(e_i) \tag{1}$$

At a given site $i$, the presence/absence of a given species $j$ follows a Bernoulli distribution parameterized with a mean probability of occurrence $\theta_{ij}$.

$$y_{ij\,|e_i,h_i} \sim Bernoulli(\theta_{ij}) \tag{2}$$

The occurrence probability of a given species $j$ varies across sites. At a given site $i$, it is modeled as the combined response (eq. 3) of the species $j$ to the observed environment $x_i$ (environment effect) and unobserved or hidden factors $h_i$ (residual effect) which represent unmeasured confounders (e.g. microbial activity) or interspecies association drivers (e.g. resource for competition, microhabitat for facilitation).

$$\begin{aligned} g(\theta_{ij}) &= \sum_{k=1}^{K} x_{ik} \cdot \beta_{kj} + \sum_{l=1}^{L} h_{il} \cdot \alpha_{lj} \\ &= \underbrace{x_i \cdot \beta_j}_{env.\ effect} + \underbrace{h_i \cdot \alpha_j}_{residual} \\ &= x_i \cdot \beta_j + z_{ij} \end{aligned} \tag{3}$$

The parameters $\beta$ and $\alpha$ represent the coefficients of species response to the environment and their latent factor loadings respectively. The link function g maps the linear response to the mean.

We assume that species are independent (eq. 4), conditional on the environmental and hidden factors.

$$y_{ij} \perp y_{ij'} \quad | x_i , h_i \tag{4}$$

We further assume that the hidden factors follow an independent normal distribution (prior) with mean vector $\mu_p$ and variance vector $\sigma_p^2$ (eq. 5).

$$h_i \sim \mathbb{N}\left(\mu_p, \sigma_p^2 \cdot I\right) \tag{5}$$

### A1.3 Inference as optimization

The goal of inference is to maximize the marginal loglikelihood (eq. 6) of observed community composition conditional to the observed environment $P(Y|E)$ w.r.t the generative parameters $\Phi = (W^e, \mathrm{A}, \mathrm{B})$, denoted as $\mathrm{p}_\Phi(Y)$.

$$\log p_\Phi (Y) = \log \int p_\Phi(Y, H)\, dH \tag{6}$$

Given that hidden factors are de facto unobserved, this requires computing an intractable integral. To circumvent this issue, we introduce a variational approximation $q(H|Y)$ of the posterior distribution of the latent variables $p(H|Y)$ within a class Q of distributions. Here, we choose the family of factorized Gaussian distributions (eq. 7).

$$Q = \left\{ q : q(H) = \prod_{i=1}^{N} \mathbb{N}\,(h_i \,;\, \mu_i \,, \sigma_i^2) \right\} \tag{7}$$

From the previous equation, the number of posterior parameters (means and variances) increases linearly with N the number of observations $O(2LN)$.

To ensure the scalability of the inference, we follow the AEVB algorithm and use a sufficiently complex parametric function $f^{ENC} : y_s \rightarrow (\mu_s, {\sigma_s}^2)$. In practice, $f^{ENC}$ is a neural network parameterized with weights $\xi$, referred to as **recognition** or **probabilistic encoder** network. The combination of the encoder and the generative model (namely, **decoder**), leads to a **variational autoencoder (VAE)** [Kingma and Welling, 2013].

This transforms the inference problem to an optimization problem where we maximize the likelihood $\mathcal{J}$ w.r.t the global parameters $\Phi$ minus the information loss $\mathcal{D}$ induced by the posterior approximation w.r.t the encoder parameters $\xi$.

$$\mathcal{J}(Y;\Phi,\xi) = \log p_{\Phi}(Y) - \mathcal{D}\left(q_{\xi}(H \mid Y) \parallel p_{\Phi}(H \mid Y)\right)$$

$$= \log p_{\Phi}(Y) - \mathbb{E}_{H\sim q(H\mid Y)} \log\left(\frac{q_{\xi}(H \mid Y)}{p_{\Phi}(H \mid Y)}\right)$$

$$= \log p_{\Phi}(Y) - \mathbb{E}_{H\sim q(H\mid Y)} \log\left(\frac{q_{\xi}(H \mid Y)\cdot p_{\Phi}(Y)}{p_{\Phi}(Y \mid H)\cdot p(H)}\right)$$

$$= \mathbb{E}_{H\sim q(H\mid Y)} \log p_{\Phi}(Y \mid H) + \mathcal{D}\left(p(H) \parallel q_{\xi}(H \mid Y)\right) \tag{8}$$

The KL divergence between the prior and approximate posterior of the latent factors, both normally distributed, can be computed in closed form. We further introduce a regularization term on the model parameters in the final optimization objective (eq. 9) to prevent overfitting.

$$\mathcal{J}_{reg}(Y;\Phi,\xi) = \underbrace{\mathbb{E}_{H\sim q(H\mid Y)} \log p_{\Phi}(Y \mid H)}_{reconstruction\ loss} + \underbrace{\mathcal{D}\left(p(H) \parallel q_{\xi}(H \mid Y)\right)}_{KL\ loss} + \underbrace{\log p(\Phi) + \log p(\xi)}_{parameter\ regularization} \tag{9}$$

The optimization objective is to minimize the total cost $J_{reg}$, which includes the **reconstruction loss** (measuring the difference between the predicted and actual species presence), the **KL divergence loss** (penalizing the divergence between the approximate and true posterior distributions of latent factors), and a **regularization** term (preventing overfitting by controlling model complexity).

In the context of this paper, the reconstruction loss is the balanced binary cross-entropy loss as the reconstruction loss, whereas we use Elasticnet regularization loss for the parameter regularization term.

## A1.4 Species association

We define the following posterior statistics over the latent factors $H$, assuming the dimension of the latent space is $L$ and the number of data samples is $N$

| | | |
|---|---|---|
| Variational mean matrix | $\mathcal{U} = (\mu_s)^N{}_{s=1}$ | $N \times L$ |
| Accumulated variance matrix | $\mathcal{S} = \sum_{s=1}^{N} diag({\sigma_s}^2)$ | $L \times L$ |
| Latent factor covariance matrix | $\hat{\Sigma} = \frac{1}{N}\sum_{s=1}^{N}(\mathcal{U}'\mathcal{U} + \mathcal{S})$ | $L \times L$ |

From there, the species-by-species residual covariance matrix can be formulated as:

$$\Sigma_r = A \cdot \hat{\Sigma} \cdot A' \qquad M \times M$$

We use the *Graphical lasso*[2] algorithm to estimate a sparse inverse for the posterior residual covariance matrix $\hat{\Omega}_r = \hat{\Sigma}_r^{-1}$ with a sample size of $N$. This corresponds to the precision matrix which encodes conditional dependencies between species in the latent space.

The corresponding partial correlation matrix encodes a Gaussian Graphical Model or Gaussian Random Field that we refer to as the species association network. It is an undirected graphical model in which the absence of a link represents pairwise independence conditional to other species.

[2] Friedman, J., Hastie, T., & Tibshirani, R. (2008). Sparse inverse covariance estimation with the graphical lasso. *Biostatistics*, *9*(3), 432-441.

## A2. Neural network architecture

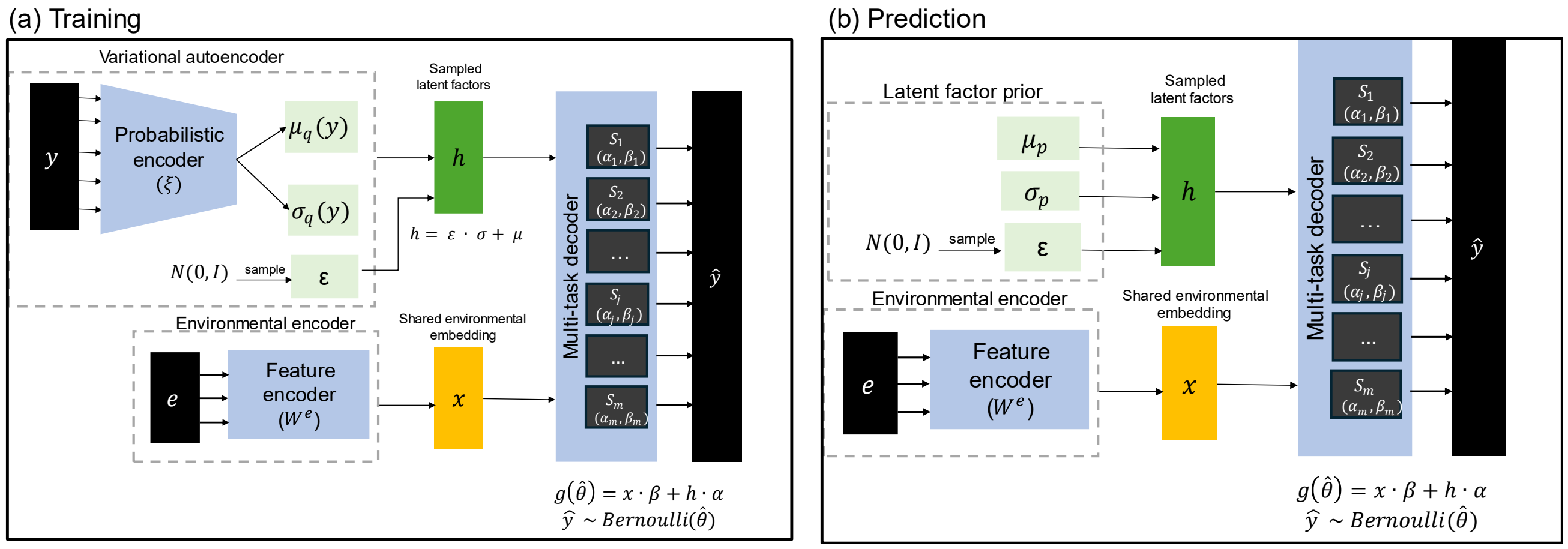

*Figure 1 - Architecture of MTEC*

The architecture of MTEC is composed of three modules:

- **A probabilistic encoder**: a fully connected neural network with parameters $\xi$ (weights and biases). During training, this network takes as input observed communities and computes the latent factors variational posterior distribution parameters: mean $\mu_q$ and variance $\sigma_q^2$. To sample from this posterior distribution, we first generate a sample $\varepsilon$ from a standard normal distribution then we compute the latent factors using the reparameterization trick (eq. 10) to preserve the gradient flow.

$$h = \varepsilon \cdot \sigma_q + \mu_q \tag{10}$$

- **An environmental feature encoder**: a fully connected neural network with parameters (weights and biases) which takes as input the raw environmental covariates and transforms them to extract a shared embedding with relevant features for all species. We use category embedding layers for categorical features and dense layers for numerical features.
- **A multi-task decoder**: a fully connected neural network that predicts multi-species probabilities of presences as a function of the shared environmental embedding and the latent factors. The latter are sampled from the posterior distribution during training on known communities and from the prior distribution on new sites.

# B. Implementation details

We fitted MTEC to our dataset of earthworm communities and environmental covariates across training sampling sites.

## B1 Model settings

### B1.1 General settings

We used a probit link g = Φ to accommodate the binary presence/absences. Given the limited size of the calibration dataset, we used a single layer in the probabilistic encoder and fixed the number of latent factors to 3 to facilitate the visualization. Furthermore, we applied a combined ridge and lasso (elastic net) regularization to the generative θ and variational ξ parameters to control model complexity.

### B1.2 Environmental feature encoder

We experimented with different settings to process environmental features both at the data preprocessing stage (dimensionality reduction) and within the feature encoder (architecture complexity).

**Feature preprocessing** We applied an one-hot-encoding to categorical features and standardized numerical and ordinal features. Due to the large number of abiotic covariates and responding taxa, we investigated different feature selection/transformation approaches: (1) End-to-end (learnt transformation); (2) preprocessing via correlation-based selection using the Variation Inflation Factor; (3) preprocessing via principal component analysis.

**Feature encoder architecture** To find the optimal configuration of network complexity and transferability amongst prediction tasks, we varied the shared feature encoder network depth, and the width (number of neurons) of each layer.

**Selection and evaluation** Combining learning architectures with the preprocessing modes yielded 6 configurations. We trained all 6 models on the full calibration dataset while varying the regularization strength in the range {1,2,5,10} · 10−4. We used a 5×2 cross-validation on the calibration dataset coupled with a paired t-test (Dietterich, 1998) to compare models in pairs. We report average performances on the calibration dataset (ROC-AUC) and the evaluation dataset (recall) as well as optimal regularization settings for each configuration in Table 5.

| **Configuration** | | | **Average predictive performances** | | **Optimal regularization** |
|---|---|---|---|---|---|
| Feature preprocessing | Feature encoder architecture | Model name | Calibration ROC-AUC | Evaluation Recall | $(\lambda_{lasso}, \lambda_{ridge}).\,10^{-4}$ |
| None | [16] | E2E-Shared16 | **0.89 ± 0.106** | **0.765** | (1,1) |
| | [32-16] | E2E-Shared32-16 | 0.861 ± 0.02 | 0.604 | (5,1) |
| PCA | [16] | PCA-Shared16 | 0.886 ± 0.02 | 0.679 | (2,1) |
| | [32-16] | PCA-Shared32-16 | 0.864 ± 0.021 | 0.574 | (10,1) |
| VIF | [16] | VIF-Shared16 | 0.864 ± 0.022 | 0.519 | (10,1) |
| | [32-16] | VIF-Shared32-16 | 0.843 ± 0.02 | 0.65 | (10,1) |

*Table 5 – Tuning combinations of environmental preprocessing and feature encoder architectures.*

We find that the end-to-end approach with a shared feature extraction layer outperformed the other strategies in terms of recall on the evaluation dataset, while scores on the training dataset were comparable (Table 5).

# B2 Training settings

## B2.1 Optimization

We fit the model using the Adam optimization algorithm for at most 400 iterations (epochs), using a batch size of 32. To prevent overfitting, we used early stopping to stop the training whenever the objective function of the validation set stopped improving after more than 10 epochs.

## B2.2 Dealing with imbalance

Like most species' occurrence data, the earthworm dataset used in this study exhibits a long-tailed species prevalence distribution where a vast majority of species occur in less than 10% of the community samples while.

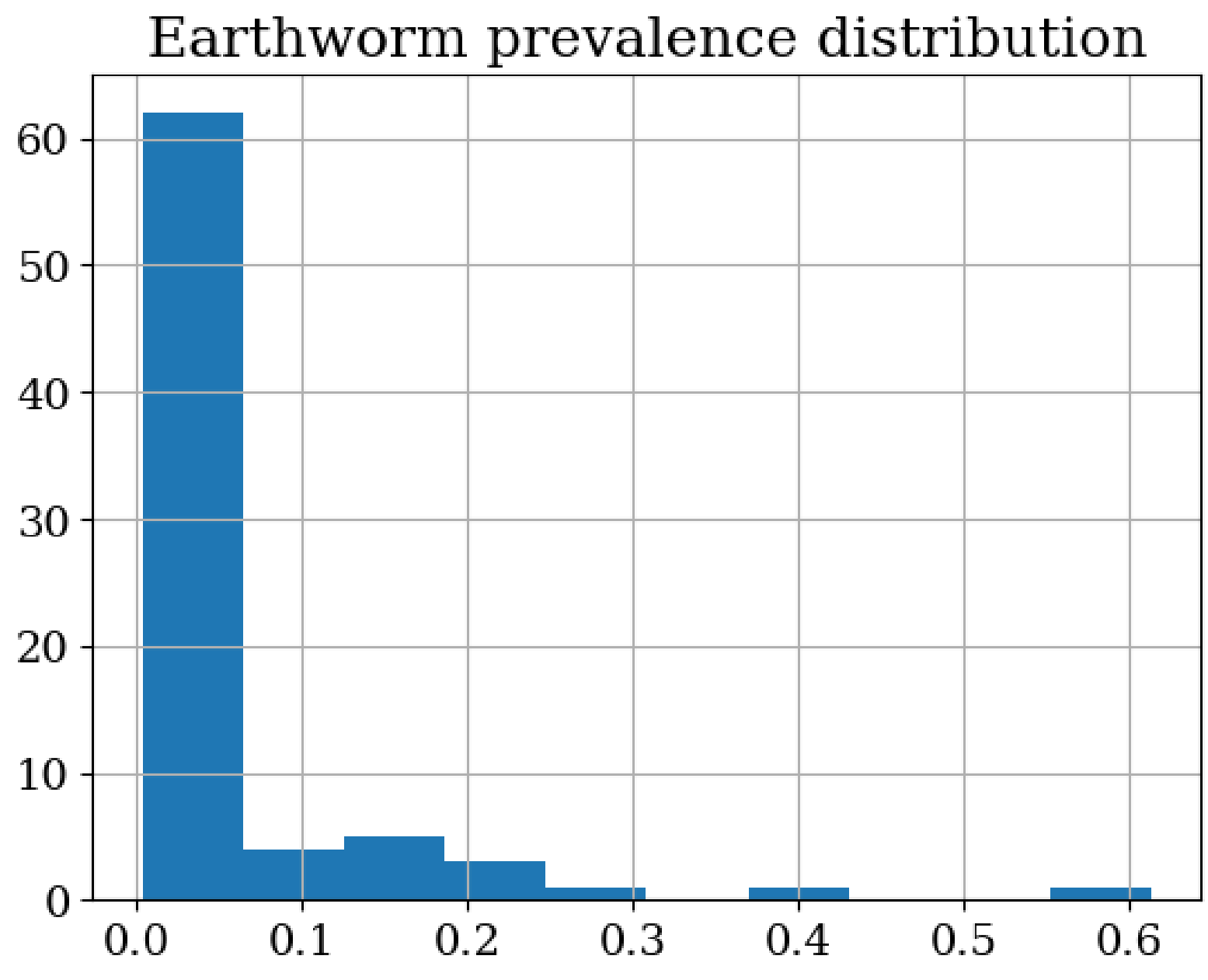

*Figure 2 - Histogram of earthworm species prevalence*

Due to the imbalance of presences and absences, random sampling of observations may result in training sets that contain only absences for rare species. To circumvent this issue and to enable a robust estimation of the abiotic niche for all taxa, we devised an Algorithm that ensures at least 5 occurrences for each species in the sampled set (min_occur = 5).

```
Input  : Y: Occurrence matrix of dimensions (N × M), min_occur: integer, tsize: training set size
Output: T, V: set of training and validation observations
/* Initialization */
T := ∅;
V := {1, 2, ..., N};
Req := REP(min_occur, M) ;                                  // Required occurrences
S := {1, 2, ..., M} ;                                       // Unsatisfied taxa
Prev := REP(0, M);
while S≠ ∅ do
    /* Satisfy rarest taxa first */
    potential := available presences for each unsatisfied taxa ;
    candidates := taxa with the smallest potential ;
    choice := select randomly from candidates ;
    nidx := Req[choice] ;
    /* Draw occurrence points */
    selected := draw nidx indices from potential[choice] ;
    T := T∩ selected ;
    V := V\ selected ;
    /* Update requirements */
    Prev := ROWSUM(Y[T, ]) ;
    Req := Req - Prev ;
    S := {1 ≤ t ≤ M | Req[t] > 0} ;
end
/* Fill rest of training set randomly */
ncomp := SIZE(T) − tsize;
if ncomp < 0 then
    selected := draw uniformly ncomp sites from V;
    T := T∩ selected ;
    V := V\ selected ;
end
```

We ensured all taxa appeared in each training set and weighted likelihood presences (positive loss) by their odds relative to absences (negative loss). To prevent bias fitting, we initialized the mean prediction intercepts to each taxon's logit-prevalence, while other parameters used the Glorot-uniform initializer (Glorot & Bengio, 2010).

# C. Supplementary results

## C1 Comparison of model predictive performances

In this section, we report species-wise predictive performances both on the calibration dataset (cross-validation) and on the evaluation dataset.

### C1.1 Statistical ranking test results

Figure 2 shows the distribution of species-wise performances on the calibration dataset for MTEC and single species SDM models (GLM, RF, GBM).

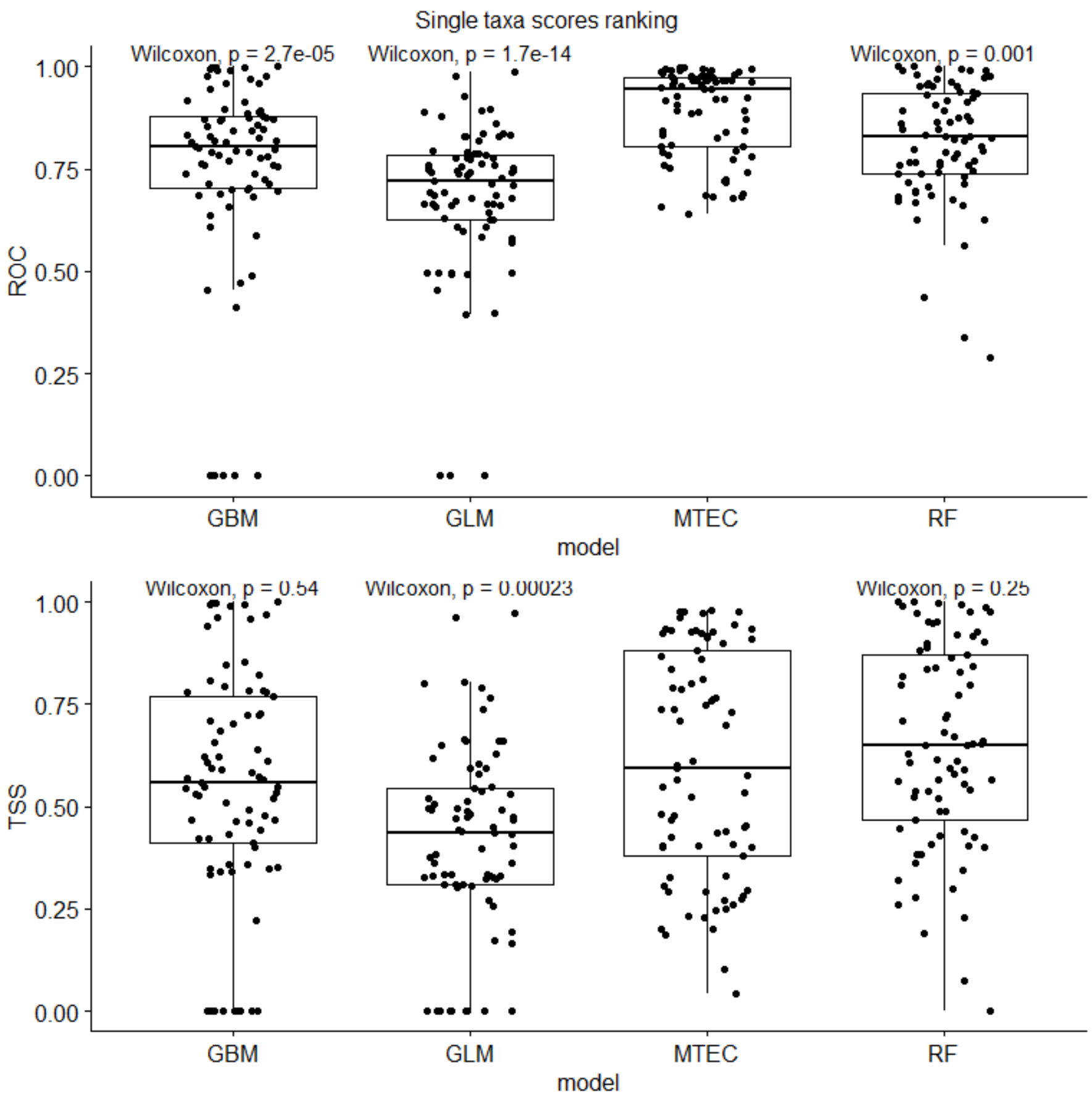

*Figure 4 - Species-level cross-validation performances for SDM models (GLM, GBM, RF) and MTEC in terms of ROC-AUC (top) and TSS (bottom).*

## C1.2 Table of species-wise predictive performances

| Target | Prevalence | threshold | True Skill Statistic | | | | ROC-AUC | | | | Evaluation recall | | | |
|---|---|---|---|---|---|---|---|---|---|---|---|---|---|---|
| | | | MTEC | GLM | GBM | RF | MTEC | GLM | GBM | RF | MTEC | GLM | GBM | RF |
| **Average** | 0.05243048 | 0.65012987 | **0.746** | 0.411 | 0.552 | 0.646 | **0.912** | **0.681** | **0.753** | **0.816** | **0.914** | **0.519** | **0.302** | **0.250** |
| **Predictive performances for each taxa** | | | | | | | | | | | | | | |
| **Abart** | 0.010 | 0.88 | **0.947** | 0.326 | 0.853 | 0.869 | **0.991** | 0.663 | 0.913 | 0.914 | **1** | | | |
| **Alchae** | 0.007 | 0.88 | 0.95 | 0.331 | **0.958** | 0.953 | **0.996** | 0.663 | 0.969 | 0.968 | **1** | **1.000** | | |
| **Alchlo** | 0.150 | 0.46 | 0.437 | **0.475** | 0.430 | 0.405 | **0.782** | 0.753 | 0.769 | 0.761 | | | | |
| **AlChlo_ChIn** | 0.055 | 0.58 | 0.333 | 0.193 | 0.222 | 0.227 | **0.707** | 0.570 | 0.587 | 0.562 | **1** | 0.165 | | |
| **AlChlo_ChLe** | 0.202 | 0.48 | 0.263 | 0.334 | **0.352** | 0.300 | 0.687 | 0.692 | **0.695** | 0.676 | **1** | | | |
| **AlChlo_ChWa** | 0.013 | 0.53 | **0.859** | 0.309 | 0.558 | 0.562 | **0.95** | 0.599 | 0.761 | 0.739 | | | | |
| **AlChlo_PoCh** | 0.005 | 0.69 | **0.871** | 0.000 | 0.000 | 0.629 | **0.97** | 0.395 | 0.000 | 0.718 | | | | |
| **AlChlo_PoIn** | 0.006 | 0.74 | **0.868** | 0.330 | 0.493 | 0.537 | **0.969** | 0.626 | 0.702 | 0.708 | 1 | | | |
| **AlChlo_PoLe** | 0.029 | 0.46 | **0.612** | 0.536 | 0.508 | 0.447 | **0.883** | 0.762 | 0.814 | 0.759 | | | | |
| **ApArve** | 0.005 | 0.86 | 0.973 | 0.663 | **0.990** | **0.990** | **0.999** | 0.830 | 0.990 | 0.990 | **1** | 0.333 | | |
| **ApBali** | 0.004 | 0.89 | **0.975** | 0.332 | 0.000 | 0.947 | **0.998** | 0.663 | 0.000 | 0.952 | **1** | | | |
| **Apcali** | 0.613 | 0.6 | 0.223 | 0.325 | **0.401** | 0.383 | 0.712 | 0.687 | **0.737** | 0.736 | 0.991 | 0.248 | **0.774** | 0.607 |
| **ApCali_Al** | 0.016 | 0.48 | 0.815 | 0.736 | 0.727 | **0.882** | 0.949 | 0.837 | 0.894 | **0.954** | | | | |
| **ApCali_Me** | 0.155 | 0.41 | 0.489 | 0.472 | **0.567** | 0.556 | 0.818 | 0.776 | **0.833** | 0.817 | **1** | | | |
| **Apcupu** | 0.045 | 0.47 | 0.41 | 0.438 | **0.532** | 0.522 | 0.791 | 0.754 | **0.805** | 0.799 | **1** | 0.067 | | |
| **Apgeor** | 0.006 | 0.84 | **0.935** | 0.164 | 0.784 | 0.652 | **0.994** | 0.579 | 0.879 | 0.804 | | | | |
| **Apgiar** | 0.121 | 0.41 | 0.505 | 0.529 | 0.534 | **0.565** | 0.828 | **0.831** | 0.818 | 0.825 | 0.946 | **0.973** | | |
| **ApGiar_Vo** | 0.008 | 0.81 | **0.93** | 0.592 | 0.703 | 0.606 | **0.985** | 0.783 | 0.842 | 0.765 | | | | |
| **Apicte** | 0.136 | 0.51 | 0.329 | 0.375 | **0.564** | **0.564** | 0.707 | 0.691 | **0.847** | 0.846 | **0.83** | 0.216 | 0.114 | |
| **ApLong_Lo** | 0.130 | 0.53 | **0.46** | 0.434 | 0.444 | 0.425 | **0.804** | 0.759 | 0.777 | 0.768 | | | | |
| **ApLong_LoAm** | 0.004 | 0.89 | **0.93** | 0.000 | 0.000 | 0.189 | **0.994** | 0.492 | 0.000 | 0.438 | 0.8 | | | |
| **ApLong_LoRi** | 0.036 | 0.47 | 0.543 | **0.545** | 0.521 | 0.519 | **0.834** | 0.787 | 0.760 | 0.740 | | | | |
| **ApLong_LoRiVi** | 0.008 | 0.64 | **0.915** | 0.790 | 0.359 | 0.648 | **0.982** | 0.892 | 0.657 | 0.833 | | | | |
| **Aprose** | 0.375 | 0.5 | 0.326 | 0.305 | **0.340** | 0.279 | **0.704** | 0.679 | 0.701 | 0.669 | 0.313 | | **0.014** | **0.129** |
| **Avlig** | 0.010 | 0.77 | **0.913** | 0.617 | 0.683 | 0.827 | **0.984** | 0.793 | 0.866 | 0.937 | **1** | | | |
| **Bieise** | 0.017 | 0.7 | **0.928** | 0.766 | 0.639 | 0.680 | **0.983** | 0.897 | 0.859 | 0.893 | | | | |
| **Bosp** | 0.008 | 0.85 | 0.962 | 0.494 | 0.941 | **0.985** | **0.996** | 0.747 | 0.977 | 0.992 | | | | |
| **Deatte** | 0.020 | 0.65 | 0.878 | 0.361 | 0.822 | **0.902** | **0.979** | 0.678 | 0.960 | 0.974 | **1** | **1.000** | 0.375 | |
| **Debybl** | 0.004 | 0.84 | **0.917** | 0.000 | 0.000 | 0.536 | **0.986** | 0.000 | 0.472 | 0.695 | **1** | | | |
| **Deocta** | 0.051 | 0.46 | **0.72** | 0.514 | 0.591 | 0.594 | **0.926** | 0.791 | 0.870 | 0.865 | **1** | 0.750 | 0.200 | |
| **Depygm** | 0.005 | 0.89 | **0.957** | 0.659 | 0.000 | 0.915 | **0.994** | 0.830 | 0.000 | 0.925 | 0.2 | | | |
| **Derubi** | 0.048 | 0.42 | **0.633** | 0.443 | 0.464 | 0.466 | **0.897** | 0.736 | 0.795 | 0.767 | **0.905** | | | |
| **DeSubr** | 0.035 | 0.51 | **0.522** | 0.269 | 0.420 | 0.409 | **0.827** | 0.644 | 0.686 | 0.687 | **1** | | | |
| **DiOmod** | 0.015 | 0.86 | **0.971** | 0.492 | 0.792 | 0.834 | **0.991** | 0.740 | 0.897 | 0.931 | | | | |
| **Dipilo** | 0.005 | 0.88 | 0.939 | 0.323 | 0.657 | **0.976** | **0.987** | 0.609 | 0.819 | 0.978 | **1** | | | |
| **Eifeti** | 0.007 | 0.68 | **0.913** | 0.000 | 0.000 | 0.073 | **0.986** | 0.496 | 0.412 | 0.339 | **1** | | | |
| **Eitetr** | 0.066 | 0.49 | 0.486 | 0.468 | 0.466 | **0.489** | **0.823** | 0.746 | 0.798 | 0.765 | **0.768** | | | |
| **Hagord** | 0.011 | 0.52 | **0.755** | 0.396 | 0.348 | 0.384 | **0.922** | 0.584 | 0.638 | 0.625 | | | | |
| **Hemoni** | 0.007 | 0.84 | **0.93** | 0.327 | 0.844 | 0.899 | **0.995** | 0.663 | 0.958 | 0.955 | | | | |
| **Heocul** | 0.010 | 0.51 | **0.878** | 0.548 | 0.477 | 0.401 | **0.961** | 0.715 | 0.725 | 0.627 | | | | |
| **Hosamn** | 0.004 | 0.88 | 0.97 | 0.661 | **1.000** | **1.000** | 0.997 | 0.829 | **1.000** | **1.000** | **1** | | | |
| **Krcal** | 0.008 | 0.65 | **0.919** | 0.000 | 0.549 | 0.717 | **0.977** | 0.000 | 0.758 | 0.831 | **1** | | | |
| **Lucast** | 0.064 | 0.51 | 0.378 | 0.451 | **0.571** | 0.541 | 0.742 | 0.775 | 0.827 | **0.829** | 0.99 | **0.848** | | |
| **LuCast_Di** | 0.263 | 0.51 | 0.242 | 0.255 | **0.341** | 0.318 | 0.663 | 0.627 | **0.690** | 0.681 | | | | |
| **Lucent** | 0.051 | 0.45 | **0.542** | 0.433 | 0.526 | 0.487 | **0.841** | 0.743 | 0.802 | 0.789 | **0.875** | | | |
| **Lufest** | 0.049 | 0.44 | **0.706** | 0.627 | 0.623 | 0.651 | **0.903** | 0.861 | 0.871 | 0.880 | | | | |
| **Lufrie** | 0.127 | 0.45 | 0.55 | 0.650 | **0.725** | 0.711 | 0.861 | 0.877 | 0.889 | **0.893** | **1** | 0.130 | | |
| **Lumeli** | 0.004 | 0.87 | **0.965** | 0.000 | 0.609 | 0.659 | **0.992** | 0.494 | 0.779 | 0.793 | **1** | | | |
| **Lurube** | 0.113 | 0.48 | 0.543 | 0.578 | **0.606** | 0.614 | 0.82 | 0.818 | 0.854 | **0.863** | 0.038 | | | |
| **LuRube_Ca** | 0.009 | 0.59 | **0.691** | 0.172 | 0.000 | 0.000 | **0.891** | 0.396 | 0.491 | 0.288 | **1** | | | |
| **Luterr** | 0.216 | 0.49 | 0.35 | **0.361** | 0.348 | 0.360 | 0.705 | 0.687 | **0.713** | 0.693 | **0.96** | 0.054 | | |

| | | | | | | | | | | | | | |
|---|---|---|---|---|---|---|---|---|---|---|---|---|---|
| **Midubi** | 0.014 | 0.59 | **0.899** | 0.381 | 0.594 | 0.772 | **0.97** | 0.656 | 0.792 | 0.876 | **0.923** | 0.077 | | |
| **Miphos** | 0.016 | 0.52 | **0.918** | 0.594 | 0.779 | 0.796 | **0.972** | 0.743 | 0.874 | 0.867 | 0.857 | | | |
| **Muminu** | 0.040 | 0.56 | **0.481** | 0.301 | 0.411 | 0.344 | **0.794** | 0.607 | 0.682 | 0.662 | **1** | | | |
| **Occomp** | 0.014 | 0.58 | **0.904** | 0.659 | 0.781 | 0.887 | **0.971** | 0.837 | 0.916 | 0.930 | **1** | | | |
| **Occyan** | 0.223 | 0.56 | 0.252 | 0.308 | **0.357** | 0.258 | 0.685 | 0.670 | **0.699** | 0.671 | **0.97** | 0.388 | | 0.015 |
| **OcCyanGr** | 0.025 | 0.44 | **0.621** | 0.488 | 0.421 | 0.439 | **0.891** | 0.777 | 0.712 | 0.712 | | | | |
| **Oclact** | 0.045 | 0.54 | 0.494 | 0.521 | 0.468 | **0.578** | 0.814 | 0.758 | 0.814 | **0.822** | 1 | | | |
| **Prampl** | 0.008 | 0.61 | **0.897** | 0.308 | 0.544 | 0.671 | **0.978** | 0.628 | 0.738 | 0.775 | 1 | | | |
| **Pranti** | 0.031 | 0.4 | 0.48 | 0.603 | 0.722 | **0.723** | 0.867 | 0.789 | 0.886 | **0.918** | **1** | | | |
| **Prbise** | 0.004 | 0.84 | 0.937 | 0.000 | **0.996** | 0.797 | 0.987 | 0.494 | **0.996** | 0.859 | **1** | | | |
| **PrFrag** | 0.009 | 0.74 | 0.926 | **0.962** | 0.707 | 0.653 | **0.986** | 0.975 | 0.830 | 0.745 | **1** | **1.000** | | |
| **Prprat** | 0.004 | 0.89 | **0.97** | 0.333 | 0.333 | 0.926 | **0.994** | 0.662 | 0.610 | 0.935 | | | | |
| **Prpyre** | 0.010 | 0.69 | **0.914** | 0.482 | 0.584 | 0.844 | **0.982** | 0.772 | 0.844 | 0.937 | | | | |
| **Prtube** | 0.006 | 0.76 | **0.925** | 0.490 | 0.549 | 0.611 | **0.988** | 0.729 | 0.756 | 0.788 | | | | |
| **Samama** | 0.107 | 0.37 | 0.392 | 0.404 | **0.461** | 0.429 | 0.783 | 0.712 | **0.791** | 0.760 | **1** | | | |
| **ScAqui** | 0.010 | 0.76 | **0.905** | 0.504 | 0.784 | 0.840 | **0.982** | 0.722 | 0.875 | 0.907 | **1** | | | |
| **ScBrev** | 0.004 | 0.87 | 0.98 | 0.331 | 0.997 | **1.000** | **1** | 0.662 | 0.997 | **1.000** | | | | |
| **Sccors** | 0.027 | 0.67 | 0.968 | 0.973 | 0.963 | **0.976** | 0.991 | 0.988 | **0.992** | **0.992** | 0.333 | **0.333** | **0.333** | |
| **ScDuge** | 0.005 | 0.85 | **0.989** | 0.000 | 0.000 | 0.953 | **0.999** | 0.495 | 0.000 | 0.960 | | | | |
| **Scmifu** | 0.004 | 0.87 | 0.953 | 0.000 | 0.994 | **0.995** | **0.995** | 0.498 | 0.994 | **0.995** | **1** | **1.000** | | |
| **ScMono_Id** | 0.006 | 0.87 | 0.958 | 0.799 | 0.969 | **0.973** | **0.993** | 0.888 | 0.976 | 0.980 | **1** | | | |
| **Scporo** | 0.004 | 0.88 | **0.951** | 0.000 | 0.621 | 0.589 | **0.981** | 0.000 | 0.782 | 0.729 | | | | |
| **Scrhod** | 0.007 | 0.84 | **0.932** | 0.473 | 0.770 | 0.920 | **0.981** | 0.733 | 0.871 | 0.951 | **1** | 0.286 | | |
| **Scsavi** | 0.031 | 0.45 | **0.883** | 0.803 | 0.809 | 0.863 | **0.974** | 0.929 | 0.945 | 0.963 | | | | |
| **Zoatla** | 0.008 | 0.83 | 0.956 | 0.495 | **0.993** | **0.993** | 0.996 | 0.747 | **0.997** | 0.995 | | | | |
| **Zograff** | 0.004 | 0.88 | **0.922** | 0.000 | 0.000 | 0.817 | **0.987** | 0.453 | 0.454 | 0.845 | **1** | **1.000** | | |

*Table 6 - Detailed evaluation metrics across models on average and for each species. The best model with respect to each metric is highlighted in bold. In red, we highlight low performing models (<0.3 TSS, <0.7 ROC-AUC).*

## C1.3 Effect of species prevalence on model performances

Here, we examine how the cross-validation predictive performances in the calibration dataset vary along a gradient of species prevalences from rarest to most prevalent.

We find that both MTEC and machine learning based SDMs exhibit generally decreasing ROC-AUC with prevalence (Fig 4), except GLMs. Furthermore, MTEC obtained better scores for rare species, likely due to imbalance correction strategies during training. Additionally, single-species SDM models struggled to converge for some species with very low prevalence (fewer than 10 records). For more prevalent species, performances were variable, and this can potentially be attributed to the difference in niche complexity between species.

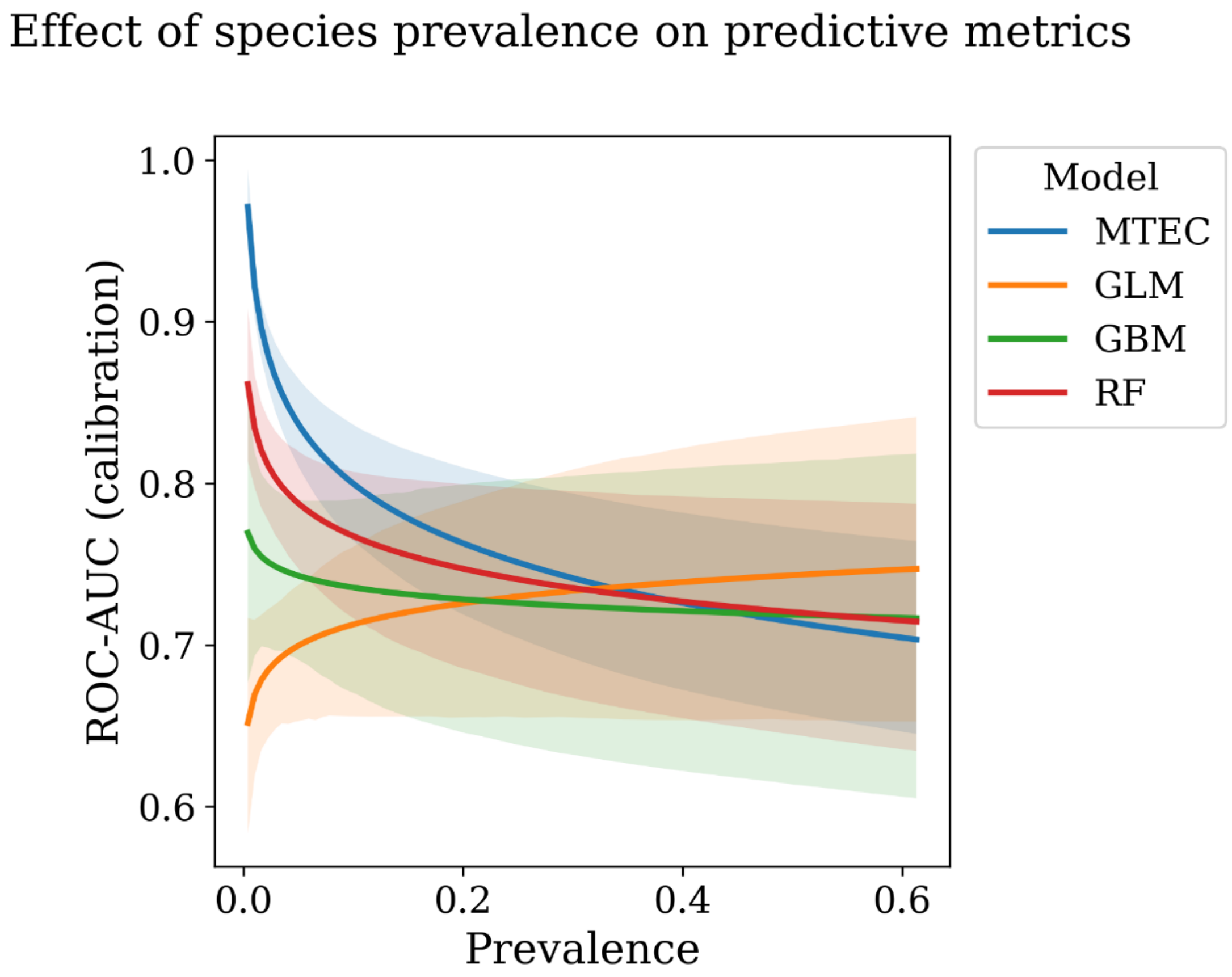

*Figure 5 - Effect of species prevalence, computed as proportion of presence observations in the calibration dataset, on the cross-validation performances (calibration dataset) in terms of ROC-AUC across models.*

## C2 Earthworm association network

In section A1.4, we described the process to infer the earthworm association network from the latent factor estimation. Figure 6 shows the resulting association network. This network illustrates associations between earthworm taxa that are not accounted for by their shared responses to environmental variables. Instead, these associations are structured by latent factors, which may reflect unmeasured environmental influences (e.g., microhabitat conditions) or biotic interactions (e.g., microbial activity).

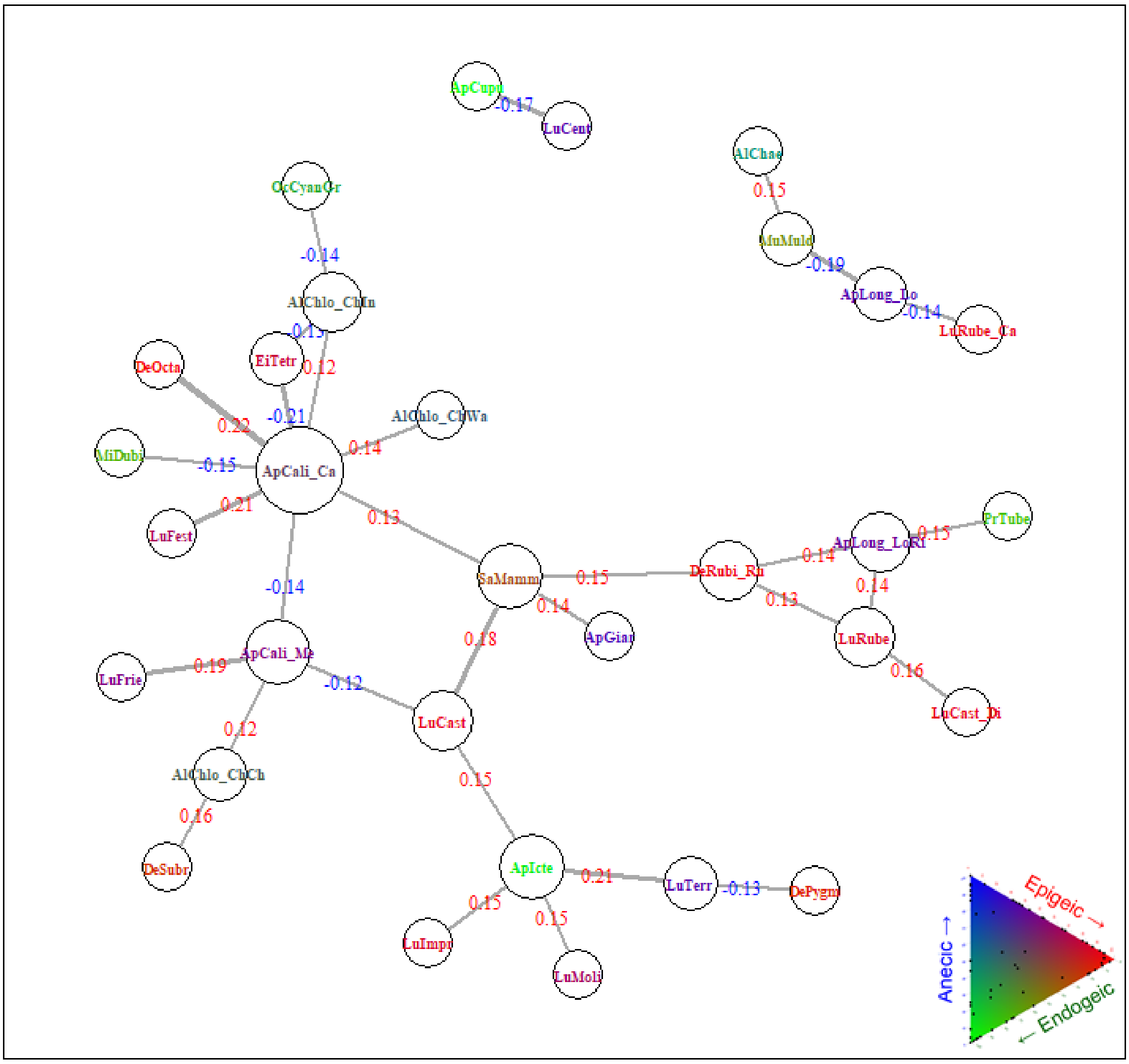

*Figure 6 - Earthworm association network. Each node (circles) represents a modeled taxon, we only represent taxa involved in significant associations. Node color corresponds to the taxa position in the ecological categories simplex (epigeic, endogeic, anecic). Edges between two taxa nodes represent significant associations. Association strength (partial correlation) normalized between -1 and 1, is shown on each node in red (resp. blue) for positive (resp.negative) associations.*

The applied method identified 31 significant pairwise associations, resulting in a network density of 0.015. These associations involved only 31 of the 77 modeled taxa, and the network visualization was accordingly restricted to these taxa. Notably, the partial correlations—representing the strength of associations and visualized as edge attributes—were relatively weak and uniform, ranging from 0.14 to 0.22 in absolute value. This suggests that, overall, earthworm distributions are primarily governed by abiotic factors.

The resulting network comprised three connected components. The largest component encompassed 80% of the nodes and 87% of the edges, and featured a hub formed by four of the most prevalent taxa: Lumbricus castaneus, Satchellius mammalis, and two subspecies of Aporrectodea caliginosa—A. caliginosa caliginosa, which is widespread, and A. caliginosa meridionalis, which is restricted to southern regions. These two subspecies were rarely co-occurring, as evidenced by their negative association.

Additionally, a positive ternary association was observed among Lumbricus rubellus, Aporrectodea longa ripicola, and Dendrobaena rubidus rubidus—three epigeic to epianecic species predominantly found in northern areas.

Most species represented in the network were classified as epigeic, epi-endogeic, or epi-anecic. This pattern may reflect the influence of an unmodeled above-ground factor, such as vegetation type, which affects litter quality (e.g., leaf carbon-to-nitrogen ratio) and its seasonality (e.g., deciduous vs. evergreen trees).

Interestingly, species positioned toward the “epigeic” end of Bouché’s ecological categories triangle (including epigeics, epi-endogeics, and epi-anecics) were predominantly involved in positive associations with endogeic and anecic species. This suggests a potential facilitative interaction, where surface-dwelling species may benefit from the use of burrows and galleries constructed by soil-dwelling taxa. These subterranean structures could serve as refuges from above-ground stressors such as cold temperatures, wind, drought, or predation by birds, mammals, or reptiles.

Notably, endemic or narrow-range species were almost entirely absent from the association network, which was predominantly composed of widespread, peregrine taxa. This pattern suggests that the success of eurytopic species in dispersing and colonizing previously inaccessible habitats may be attributed not only to their inherent competitive and dispersal capacities [Mathieu and Jonathan Davies, 2014][3] but also to positive interspecific interactions, as evidenced by the observed association patterns.

[3] Mathieu, J., & Jonathan Davies, T. (2014). Glaciation as an historical filter of below-ground biodiversity. *Journal of Biogeography*, *41*(6), 1204-1214.

# C3 Limiting factors maps

We computed the SHAP values for all environmental covariates for each species across samples of the calibration dataset. The absolute value of the SHAP statistic reflects the local importance of an environmental factor at a specific site. Mapping these values geographically revealed spatial variability in the effect of the environmental factors. We provide in the supplementary file LimitingFactorMaps.zip all limiting factor maps for modeled taxa.

# C4 Earthworm response groups

Pairwise distances among species based on SHAP values were calculated, followed by hierarchical clustering (Ward, 1963) to unravel response groups for each feature group (precipitation, land cover and soil physico-chemical properties). The optimal number of clusters was identified using the GAP statistic (Tibshirani et al., 2001).

## C4.1 SHAP based precipitation response groups

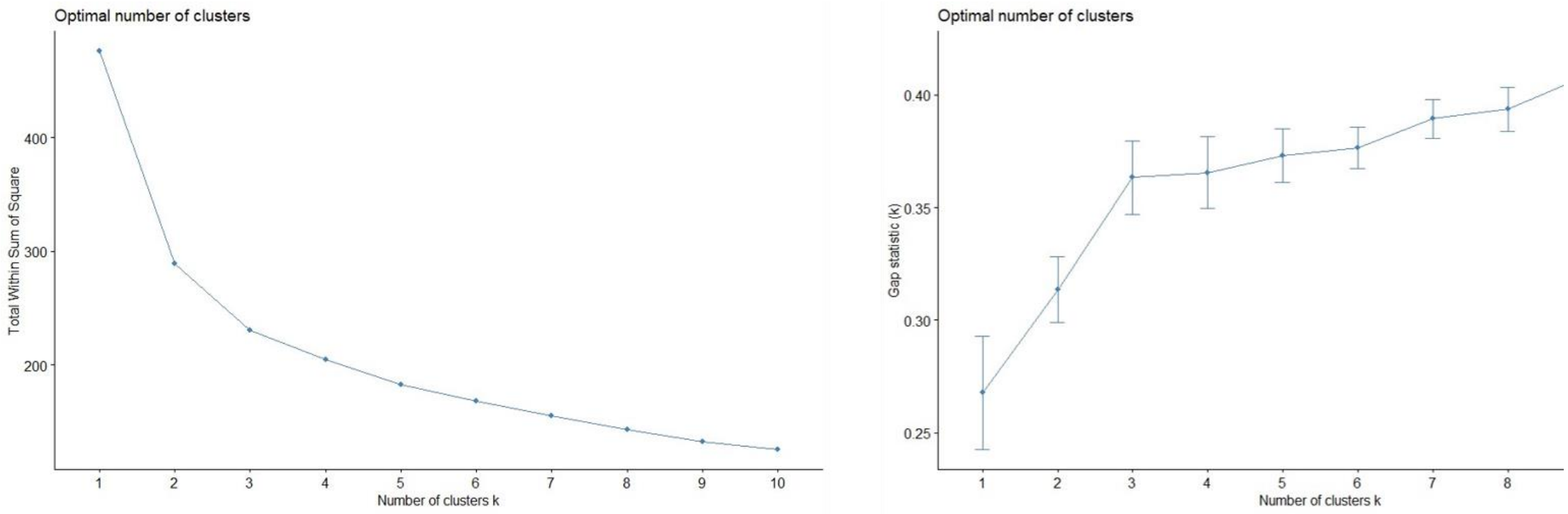

*Figure 7 – Selection of the number of precipitation response groups.*

## C4.2 SHAP based land cover response groups

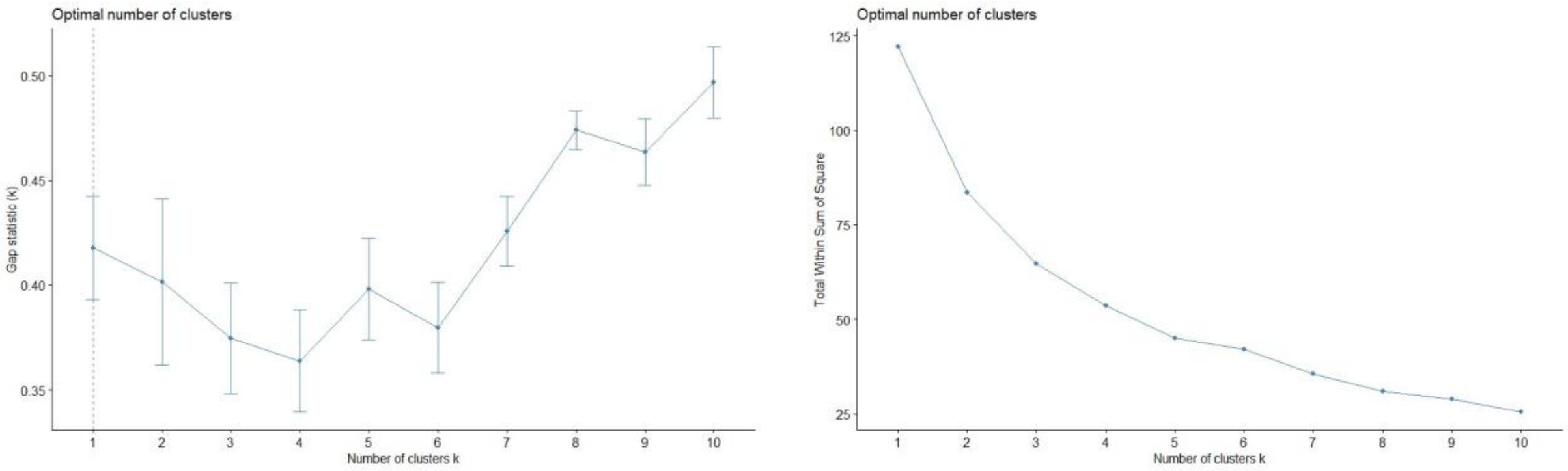

*Figure 8 - Selection of the number of land cover response groups*

## C4.3 SHAP based temperature response groups

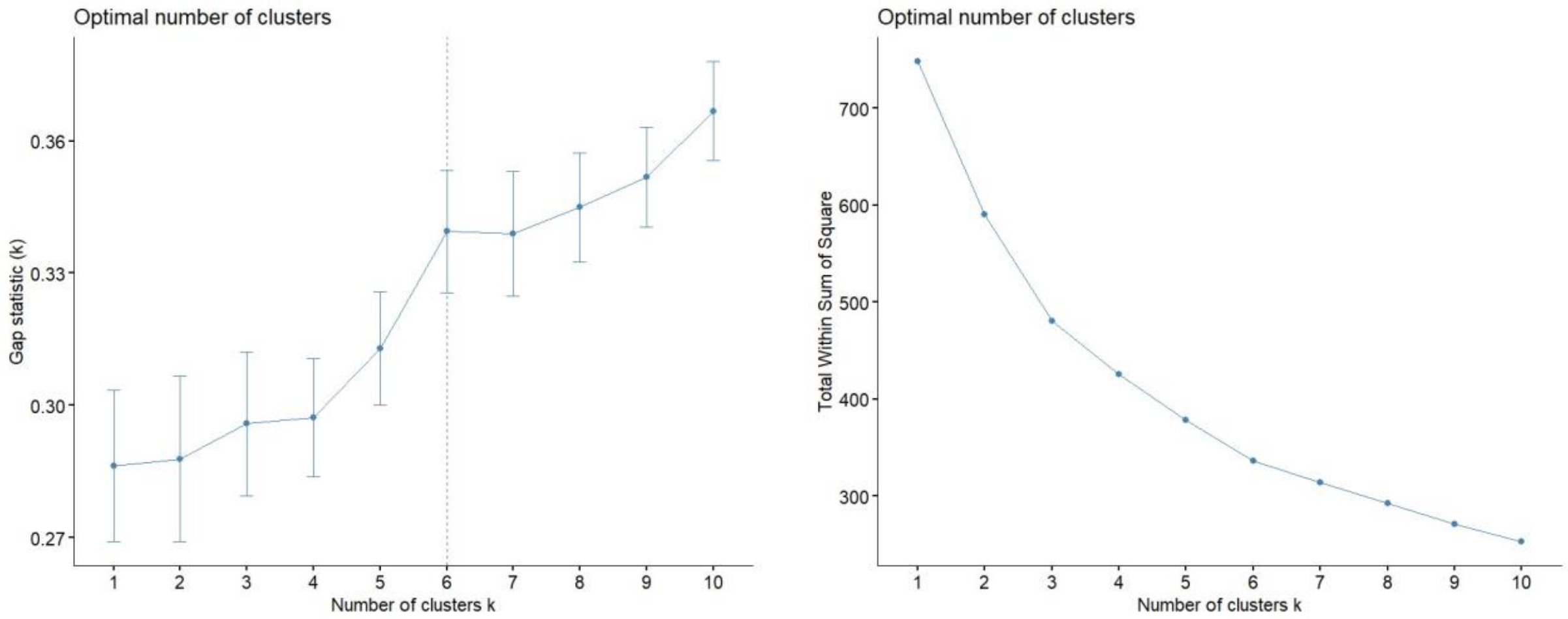

*Figure 9 - Selection of the number of temperature response groups*

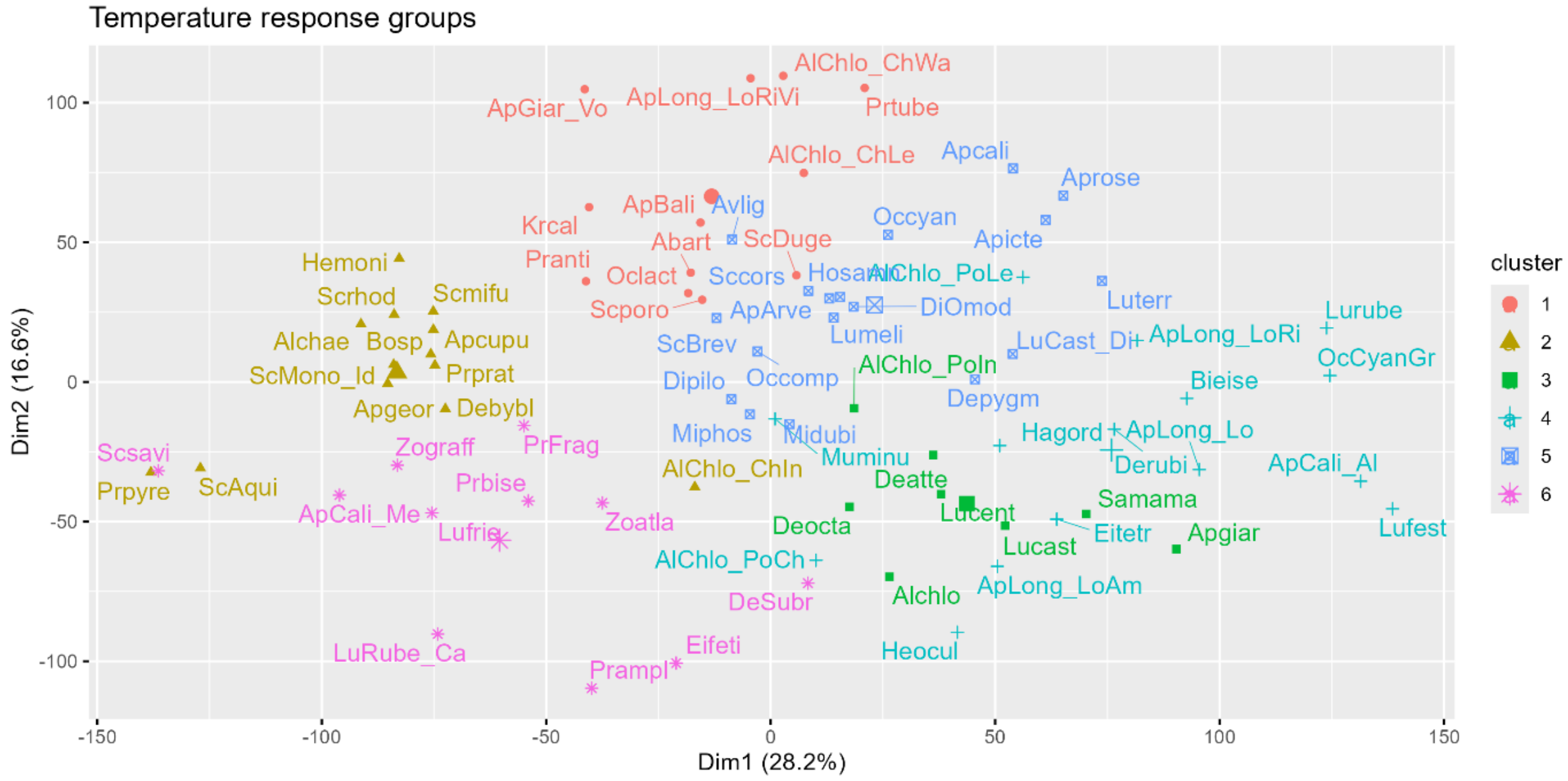

*Figure 10 - Temperature response groups*

## C4.4 SHAP based soil response groups

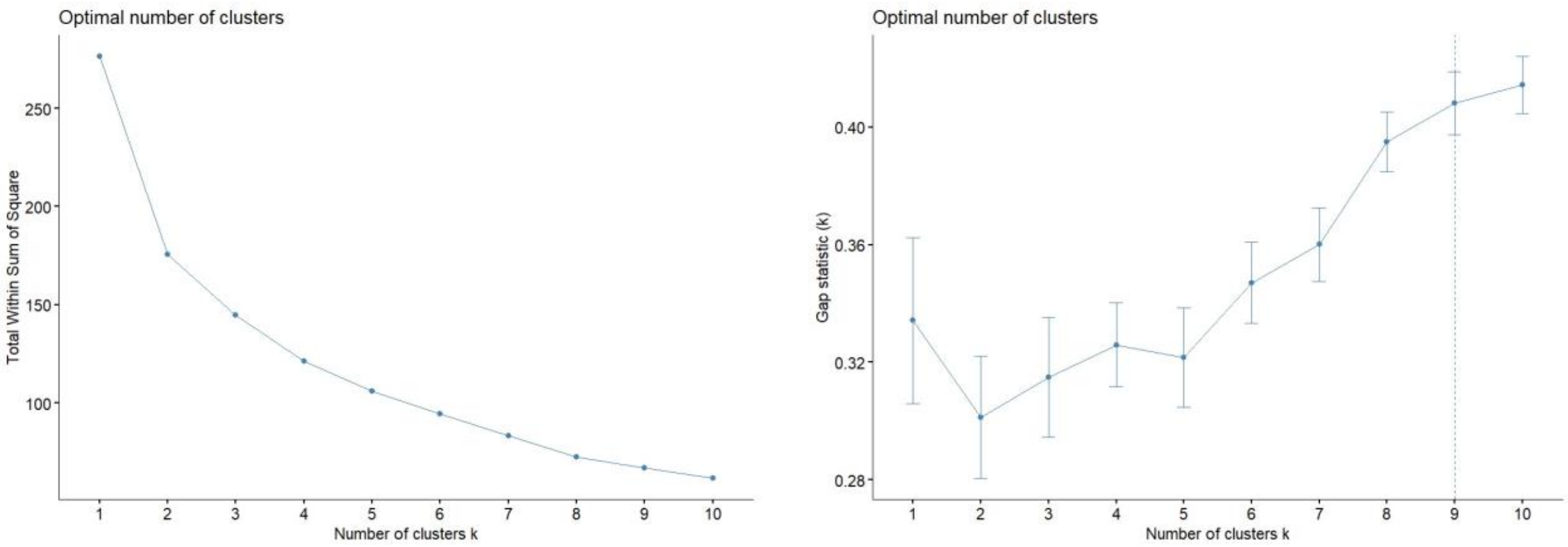

*Figure 11 - Selection of the number of soil response groups*

We analyzed the response shapes of earthworm taxa to soil acidity and organic matter quality (measured by the C:N ratio) and quantity (organic carbone), the top edaphic properties exploited by the model. The SHAP multi-species summary plots are illustrated in Figure 12.

Earthworms have different preferences in terms of organic matter quality [Curry and Schmidt, 2007]. In tropic ecosystems, endogeics are further classified depending on the quality of the consumed organic matter, from the richest to the poorest: polyhumic, mesohumic and oligohumic [Lavelle, 1988]. In our study, the gradient from “acid soils with rich SOM rich” to “alkaline soil with poor SOM” highlights a trade-off between tolerance to acidity and ability to degrade the carbon.

The first two components of the PCA of the responses to pH and C:N ratio explained ∼70% of the variation and ordinated species according to tolerance to soil acidity and organic matter quality respectively. At one end, we find the typical forest assemblages (cluster 3). They include epigeics and epi-endogeics that survive in spite of acidic soils by taking advantage of the rich organic matter in the litter. In this group, we also found endogeics of the *Zophoscolex* genus, described by Qiu and Bouché [1998] as acid-tolerant. At the other end are taxa that prefer neutral, slightly alkaline environments (cluster 1) despite poorer organic matter.

An intermediate group (cluster 2) is made of species which showed little sensitivity to pH values (SHAP values closed to zero), or with preference for the average pH value. Riparian taxa stand out in a distinct cluster (cluster 4) associated to neutral wetland soils.

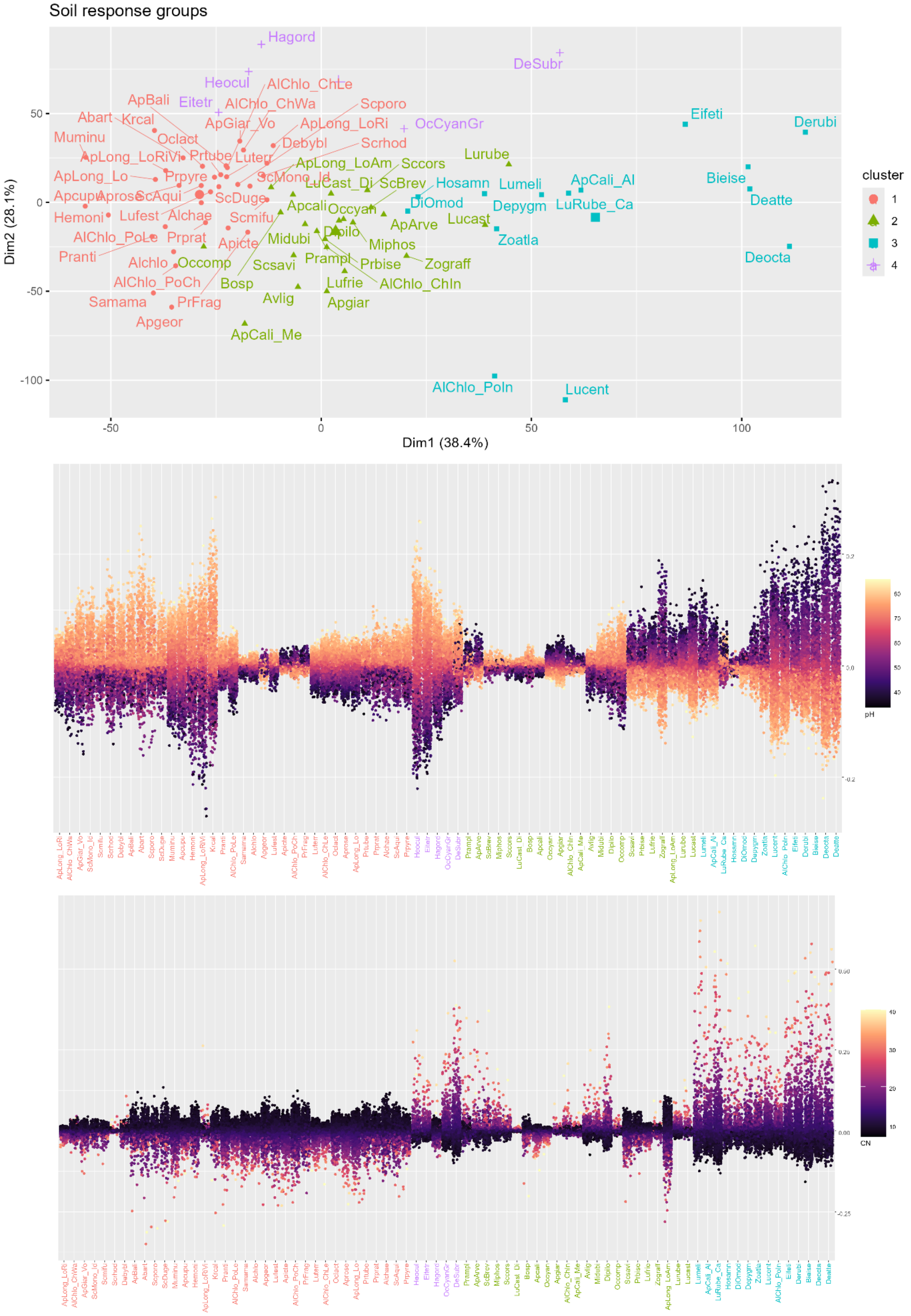

*Figure 12 – Earthworm response to soil physico-chemical properties. (a) Soil response groups. (b) Multi-species SHAP summary plot for soil water pH. (c) Multi-species SHAP summary plot for soil organic carbon to nitrogen content ratio.*